\documentclass[11pt,titlepage,a4paper]{article}
\usepackage{bozhomac}	
\usepackage{bozhlogo}	
\usepackage{cite}

\title{\bfseries	\vspace*{-1.678902345in}
{\huge Fibre bundle formulation of \\[0.22ex] relativistic quantum mechanics}
}

\vspace{1.7ex}

\author{
Bozhidar Z. Iliev
\thanks{Department Mathematical Modeling,
Institute for Nuclear Research and \mbox{Nuclear} Energy,
Bulgarian Academy of Sciences,
Boul. Tzarigradsko chauss\'ee~72, 1784 Sofia, Bulgaria}
\thanks{E-mail address: bozho@inrne.bas.bg}
\thanks{URL: http://theo.inrne.bas.bg/$\sim$bozho/}
}

\date{
 \vspace{2.27ex}\ShortTitle{Bundle relativistic quantum mechanics}\\[0.27ex]
 \vspace{3.27ex}
\small
	\begin{tabular}{r@{$\colon\to~$}l}
 \vspace{0.09ex} Basic ideas	& November 1997, January 1998	\\[0.09ex]
 \vspace{0.09ex} Began		& February 16, 1998		\\[0.09ex]
 \vspace{0.09ex} Ended		&  March 27, 1998		\\[0.09ex]
 \vspace{0.09ex} Initial typeset& April 1--14, 1998	\\[0.09ex]
 \vspace{0.09ex} Revised	& August 1999, March 2002	\\[0.09ex]
  \vspace{0.09ex} Last updated	& March 30, 2002		\\[0.09ex]
  \vspace{0.27ex} Produced	& \fbox{\today}		\\[0.27ex]
	\end{tabular} \\[1.27ex]
\normalsize
	\begin{tabular}{r@{$\colon~$}l}
\vspace{0.27ex} LANL arXive server E-print No. & quant-ph/0201085
 						\\[0.27ex]
	\end{tabular} \\[-0.27ex]
 \vspace{4.27ex}{\Huge\BOZHO}	\\[4.27ex]
 \vspace{0.27ex}\Subject{Relativistic quantum mechanics,
			  Differential geometry}		\\[2.27ex]
	\begin{tabular}{r@{\hspace{0.512em}}|@{\hspace{0.512em}}l}
 \vspace{0.27ex}\MSC[2000]{81Q99, 81S99\\{}}		
&
 \vspace{0.27ex}\PACS[2001]{02.40.Ma, 02.40.Yy\\ 02.90.+p, 03.65.Pm}
	\end{tabular} \\[1.27ex]
 \vspace{0.27ex}\KeyWords{Relativistic quantum mechanics, Fibre bundles,\\
			Geometrization of relativistic quantum mechanics,\\
	Relativistic wave equations, Dirac equation, Klein-Gordon equation
			  }	\\[0.27ex]
}

\newcommand{\fibre}{\mathcal{F}} 
\newcommand{\bundle}{(\bspace,\pr,\bbase)}	
	\newcommand{\bspace}{\mathnormal{F}}	
	\renewcommand{\pr}{\pi}			
	\newcommand{\bbase}{\mathnormal{M}}	
\newcommand{\fibreover}[1]{{\bspace_{#1}}} 

\newcommand{\Hil}{\mathcal{F}}	
\newcommand{\HilB}{(\bHil,\proj,\base)}	
	\newcommand{\bHil}{\mathit{F}}	
	\newcommand{\proj}{\pi}		
	\newcommand{\base}{\mathit{M}}	

\newcommand{\Ham}{\mathcal{H}}	
\newcommand{\bHam}{\mathit{H}}	

\newcommand{\mbHam}{\Mat{\bHam}^\mathbf{m}} 

\newcommand{\dyn}[1]{\pmb{\mathbb{#1}}}	
	\newcommand{\ope}[1]{\mathcal{#1}}		 
	\newcommand{\mor}[1]{\mathit{#1}}		 
	\newcommand{\mmor}[1]{\Mat{\mathit{#1}}}	 

\providecommand{\iu}{\mathrm{i}} 

\begin{document}		

\renewcommand{\thefootnote}{\fnsymbol{footnote}} 
\maketitle				
\renewcommand{\thefootnote}{\arabic{footnote}}   

\tableofcontents		


\begin{abstract}

	We propose a fibre bundle formulation of the mathematical base of
relativistic quantum mechanics. At the present stage the bundle form of the
theory is equivalent to its conventional one, but it admits new types of
generalizations in different directions. In the bundle description the
wavefunctions are replaced with (state) sections (covariant approach) or
liftings of paths (equivalently: sections along paths) (time-dependent
approach) of a suitably chosen vector bundle over space-time whose (standard)
fibre is the space of the wavefunctions. Now the quantum evolution is
described as a linear transportation of the state sections/liftings in
the (total) bundle space. The equations of these transportations turn to be
the bundle versions of the corresponding relativistic wave equations.
Connections between the (retarded) Green functions of these equations and the
evolution operators and transports are found. Especially the Dirac and
Klein-Gordon equations are considered.

\end{abstract}

\section {Introduction}
\label{Introduction}

	This investigation is devoted to a pure geometric (re)formulation of
the (mathematical) base of relativistic quantum mechanics in terms of fibre
bundles. It can be regarded as a direct continuation
of~\cite{bp-BQM-preliminary,
	bp-BQM-introduction+transport,bp-BQM-equations+observables,
	bp-BQM-pictures+integrals,bp-BQM-mixed_states+curvature,
	bp-BQM-interpretation+discussion,
	bp-BQM-full}
where a full self-consistent fibre bundle version of non-relativistic quantum
mechanics is elaborated. Many ideas of these works can be transferred to the
relativistic case but, as we shall see below, they are not enough for the
above aim. For example, the right mathematical concept reflecting the
peculiarities of relativistic quantum evolution is the (linear) transport
along the identity map of the space-time, on the contrary to a (linear)
transport along paths (in space-time) in the non-relativistic case. A new
problem is the fact that some relativistic wave equations are linear partial
differential equations of second order; so for them the methods developed
for the treatment of Schr\"odinger equation can not be applied directly.

	The presented in~\cite[sec.~4.1]{bp-BQM-introduction+transport}
motivations for applying the theory of fibre bundles to non\ndash
relativistic quantum mechanics can \emph{mutatis mutandis} be transferred in
the case of relativistic quantum mechanics. For the purpose, fist of all, one
has to replace the 3\ndash dimensional coordinate (or other) space model with
the Minkowski spacetime. More specific changes are connected with particular
relativistic wave equations one investigates.

	In the bundle description the relativistic wavefunctions are replaced
by (state) liftings of paths, sections or sections along paths of a suitable
vector bundle with the space-time as a base and whose (standard, typical)
fibre is the space where the corresponding conventional wavefunctions (or
some combinations of them and their partial derivatives) `live'. The
operators acting on this space are replaced with appropriate lifting
s of paths or morphisms along paths and the evolution of the wavefunctions
is now described as a linear transport of the state liftings/sections
(possibly along paths) in this bundle.

	The lay-out of the paper is the following.

	A review of the bundle approach to non-relativistic quantum
mechanics is presented in Sect.~\ref{Sect2}. It contains certain basic ideas
and equations
of~\cite{bp-BQM-introduction+transport,bp-BQM-equations+observables,
	bp-BQM-pictures+integrals,bp-BQM-mixed_states+curvature,
	bp-BQM-interpretation+discussion,bp-BQM-full}
required as a starting point for the present work.

	Sect.~\ref{Sect3} is devoted to the time-dependent or Hamiltonian
bundle description of relativistic wave equations. This approach is a
straightforward generalization of the bundle description of nonrelativistic
quantum mechanics to relativistic one.
	In Subsect.~\ref{Subsect3.3} is developed a general scheme for fibre
bundle treatment of Schr\"odinger-type partial differential equations. The
method is also applicable to linear partial differential equations of higher
orders.  This is archived by transforming such an equation to a system of
first-order linear partial differential equations (with respect to a new
function) which, when written in a matrix form, is just the
Schr\"odinger-like presentation of the initial equation. Of course, this
procedure is not unique and its concrete realization depends on the physical
problem under exploration.
	In Subsect.~\ref{Subsect3.4} is given a bundle description of the
Dirac equation. Now the corresponding vector bundle over the space-time has
as a (standard) fibre the space of 4-spinors and may be called
4-spinor bundle. Since the Dirac equation is of first order, it can be
rewritten in Schr\"odinger-type form. By this reason the formalism outlined
in Sect.~\ref{Sect2} is applied to it \emph{mutatis mutandis}. In particular,
here the state of a Dirac particle is described by a lifting of paths or
section along paths of the 4-spinor bundle and equivalently rewrite Dirac
equation as an equation for linear transportation of this section with
respect to the corresponding \emph{ Dirac evolution transport} in this
bundle.
	Subsect.~\ref{Subsect3.5} contains several procedures for transforming
Klein\ndash Gordon equation to Schr\"odinger\ndash like one. After such a
presentation is chosen, we can, analogously to Dirac case in
Subsect.~\ref{Subsect3.4}, apply to it the methods of Sect.~\ref{Sect2} and
Subsect.~\ref{Subsect3.3}.  Comments on fibre bundle description of other
relativistic wave equations are given in Subsect.~\ref{Subsect3.6}.

	With Sect.~\ref{Sect4} we begin to develop the covariant approach to
bundle description of relativistic quantum mechanics.
Subsect.~\ref{Subsect4.7} contains a covariant application of the
ideas of Sect.~\ref{Sect2} to Dirac equation. The bundle, where Dirac
particles `live', is a vector bundle over space-time with the space of
4-spinors as a fibre; so here we work again with the 4-spinor bundle
of Subsect.~\ref{Subsect3.4}, but now the evolution of a Dirac particle is
described via a \emph{geometric transport} which is a
\emph{linear transport along the identity map of space-time}.
The state of a Dirac particle is represented by a section (not along paths!)
of the 4-spinor bundle and is (linearly) transported by means of the
transport mentioned. The Dirac equation itself is transformed into a
covariant Schr\"odinger-like equation.
	In Subsect.~\ref{Subsect4.8} we apply the covariant bundle approach to
Klein-Gordon equation. For this purpose we present a 5-dimensional
representation of this equation as a first-order Dirac-like equation to
which the theory of Subsect.~\ref{Subsect4.7} can be transferred
\emph{mutatis mutandis}.

	The goal of Sect.~\ref{Sect9} is to be revealed some connections
between the retarded Green functions ($\equiv$propagators) of the
relativistic wave equations and the corresponding to them evolution operators
and transports.  Generally speaking, the evolution operators (resp.\
transports) admit representation as integral operators, the kernel of which
is connected in a simple manner with the retarded Green function (resp.\
Green morphism of a bundle).  Subsect.~\ref{Subsect9.0} contains a brief
general consideration of the Green functions and their connection with the
evolution transports, if any. In Subsect.~\ref{Subsect9.1}, \ref{Subsect9.2},
and~\ref{Subsect9.3} we derive the relations mentions for Schr\"odinger,
Dirac, and Klein-Gordon equations, respectively.

	Sect.~\ref{Conclusion} closes the paper with a brief summary of the
main ideas underlying the bundle description of relativistic quantum
mechanics.

	\ref{AppendixA} contains some mathematical results
concerning the theory of (linear) transports along maps required for the
present investigation.

	In~\ref{AppendixB} are given certain formulae concerning matrix
operators, \ie matrices with operator entries, which arise naturally in
relativistic quantum mechanics.

\section
[Bundle nonrelativistic quantum mechanics (review)]
{Bundle nonrelativistic quantum mechanics\\ (review)}
\label{Sect2}

	In the series of
papers~\cite{bp-BQM-introduction+transport,bp-BQM-equations+observables,
      bp-BQM-pictures+integrals,bp-BQM-mixed_states+curvature,
      bp-BQM-interpretation+discussion,bp-BQM-full}
we have reformulated nonrelativistic quantum mechanics in terms of fibre
bundles. The mathematical base for this was the Schr\"odinger equation
	\begin{equation}	\label{2.1}
\ih\frac{\ordinary\psi(t)}{\ordinary t} = \Ham(t) \psi(t),
	\end{equation}
where $\iu\in\mathbb{C}$ is the imaginary unit, $\hbar(=h/2\pi)$ is the Planck
constant divided by $2\pi$, $\psi$ is the system's state vector belonging to
an appropriate Hilbert space $\Hil$, and $\Ham$ is the system's Hamiltonian.
Here $t$ is the time considered as an independent variable (or parameter). If
$\psi$ is known for some initial moment $t_0$, the solution of~\ref{2.1} can
be written as
	\begin{equation}	\label{2.2}
\psi(t) = \ope{U}(t,t_0)\psi(t_0)
	\end{equation}
where $\ope{U}$ is the \emph{evolution operator} of the
system~\cite[chapter~IV, sect.~3.2]{Prugovecki-QMinHS} (for details
see~\cite{bp-BQM-introduction+transport}).

	In the bundle approach the system's Hilbert space $\Hil$ is replace
with a Hilbert bundle $\HilB$ with (total, fibre) bundle space $\bHil$,
projection $\proj$, base $\base$, isomorphic fibres
$\bHil_x:=\proj^{-1}(x)$, $x\in\base$, and (standard, typical) fibre
coinciding with $\Hil$. So, there exist isomorphisms
$l_x\colon\bHil_x\to\Hil$, $x\in\base$.
	In the present work the base $\base$ will be identified with the
Minkowski space-time $M^4$ of special relativity.%
\footnote{%
The bundle formulation of nonrelativistic quantum mechanics is insensitive to
the/this choice of $\base$~\cite{bp-BQM-interpretation+discussion}; in it is
more natural to identify $\base$ with the 3-dimensional Newtonian space
$\mathbb{E}^3$ of classical mechanics.%
}

	In the \emph{Hilbert bundle description} a state vector $\psi$ and
the Hamiltonian $\Ham$ are replaced respectively by a state section along
paths $\Psi\colon\gamma\to\Psi_\gamma$ and the \emph{bundle Hamiltonian}
(morphism along paths) $\bHam\colon\gamma\to\bHam_\gamma$, given
by~\cite{bp-BQM-introduction+transport,bp-BQM-equations+observables,
      bp-BQM-interpretation+discussion}:
	\begin{equation}	\label{2.3}
\Psi_\gamma\colon t\to \Psi_\gamma(t) = l_{\gamma(t)}^{-1}\bigl(\psi(t)\bigr),
\qquad
\bHam_\gamma\colon t\to
\bHam_\gamma(t) = l_{\gamma(t)}^{-1} \circ \Ham(t) \circ l_{\gamma(t)}
	\end{equation}
where $\gamma\colon J\to\base$, $J$ being an $\mathbb{R}$-interval, is the
world line (path) of some (point-like) observer, $t\in J$, and $\circ$
denotes composition of maps.

	The bundle analogue of the evolution operator $\ope{U}$ is the
\emph{evolution transport} $\mor{U}$ along paths, both connected by
	\begin{equation}	\label{2.4}
\mor{U}_\gamma(t,s)
=l_{\gamma(t)}^{-1}\circ \ope{U}(t,s) \circ l_{\gamma(s)}
\colon\bHil_{\gamma(s)}\to\bHil_{\gamma(t)},
\qquad s,t\in J,
	\end{equation}
which governs the evolution of state liftings via (cf.~\eref{2.2})
	\begin{equation}	\label{2.5}
\Psi_\gamma(t) = \mor{U}_\gamma(t,s)\Psi_\gamma(s), \qquad s,t\in J.
	\end{equation}

	The bundle version of~\eref{2.1}, the so-called
\emph{bundle Schr\"odinger equation},
is~\cite{bp-BQM-equations+observables,bp-BQM-full}
	\begin{equation}	\label{2.6}
\mor{D}\Psi = 0 .
	\end{equation}
Here $\mor{D}$ is the derivation
along paths corresponding to $\mor{U}$, viz (cf.~\eref{LT.8}
and~\cite[definition~4.1]{bp-LTP-general}; see also~\cite{bp-normalF-LTP}
and~\cite[definition~3.4]{bp-BQM-introduction+transport})
	\begin{equation*}
	D\colon\PLift^1(E,\pi,B) \to \PLift^0(E,\pi,B)
	\end{equation*}
where $\PLift^k\HilB$ is the set of $C^k$ liftings of paths from $\base$ to
$\bHil$, and its action on a lifting $\lambda\in\PLift^1\HilB$ with
$\lambda\colon\gamma\mapsto\lambda_\gamma$ is given via
	\begin{equation}	\label{2.7}
\mor{D}_{s}^{\gamma}\lambda :=
  \lim_{\varepsilon\to 0}
\left\{  \frac{1}{\varepsilon}
 \bigl[
\mor{U}_\gamma(s,s+\varepsilon)\lambda_\gamma(s+\varepsilon)
	- \lambda_\gamma(s)
 \bigr]
\right\}
	\end{equation}
where
$D_{s}^{\gamma}(\lambda):=((D\lambda)(\gamma))(s)=(D\lambda)_\gamma(s)$.

	If $\{e_a(\gamma(s))\}$, $s\in J$ is a basis in
$\fibreover{\gamma(s)}$, the explicit action of $\mor{D}$ is
(cf.~\eref{LT.11}; see also~\cite[proposition~4.2]{bp-LTP-general}
and~\cite{bp-normalF-LTP})
	\begin{equation}	\label{2.8}
\mor{D}_{s}^{\gamma}\lambda =
\left(
\frac{\od\lambda^a_\gamma(s)}{\od s} +
\Gamma_{{\ }b}^{a}(s;\gamma)\lambda^b_\gamma(s)
\right)
e_a(\gamma(s)) .
	\end{equation}
Here the
\emph{coefficients}
$\Gamma_{{\ }a}^{b}(s;\gamma)$ of $\mor{U}$ are defined by
(cf.and~\eref{LT.12})
	\begin{equation}	\label{2.9}
\Gamma_{{\ }a}^{b}(s;\gamma) :=
\left.
\frac{\partial\left(\mor{U}_\gamma(s,t)\right)_{{\ }a}^{b}}{\partial t}
\right|_{t=s} =
- \left.
\frac{\partial\left(\mor{U}_\gamma(t,s)\right)_{{\ }a}^{b}}{\partial t}
\right|_{t=s}
	\end{equation}
where $\left(\mor{U}_\gamma(s,t)\right)_{{\ }a}^{b}$ are given via
\(
U(t,s)e_a(\gamma(s)) =:
	\sum_{b} \left(\mor{U}_\gamma(s,t)\right)_{{\ }a}^{b} e_b(\gamma(t))
\)
and are the local components of $U$ in $\{e_a\}$.

	There is a bijective correspondence between $\mor{D}$ and the
(bundle) Hamiltonian expressed by%
\footnote{%
We denote the matrix corresponding to some quantity, e.g.\ vector or
operator, in a given field of bases with the same (kernel) symbol but in
\textbf{boldface}; e.g.\
\(
\Mat{U}_\gamma(t,s)
:=
\bigl[ \left(\mor{U}_\gamma(s,t)\right)_{{\ }a}^{b} \bigr] .
\)%
}
	\begin{equation}	\label{2.10}
\Mat{\Gamma}_\gamma(t) :=
\bigl[ \Gamma_{{\ }a}^{b}(t;\gamma) \bigr] =
- \iih \mbHam_{\gamma}(t)
	\end{equation}
with
	\begin{equation*}
\mbHam_{\gamma}(t) =
\ih
\frac{\partial\mmor{U}_\gamma(t,t_0)}{\partial t}
\mmor{U}_\gamma^{-1}(t,t_0) =
\frac{\partial\mmor{U}_\gamma(t,t_0)}{\partial t}
\mmor{U}_\gamma^{}(t_0,t).
	\end{equation*}
being the \emph{matrix-bundle Hamiltonian} (for details
see~\cite{bp-BQM-equations+observables})%
\footnote{%
Note, the constant $\ih$ in~\eref{2.10} comes from the same constant
in~\eref{2.1}.%
}

	In the Hilbert space description of quantum mechanics to a dynamical
variable $\dyn{A}$ corresponds an observable $\ope{A}(t)$ which is a linear
Hermitian operator in $\Hil$. In the Hilbert bundle description to  $\dyn{A}$
corresponds a Hermitian lifting $\mor{A}$ of paths whose
restriction on $\bHil_{\gamma(t)}$ is
	\begin{equation}	\label{2.11}
\mor{A}_{\gamma}(t) =
l_{\gamma(t)}^{-1} \circ \ope{A}(t) \circ l_{\gamma(t)}
\colon  \bHil_{\gamma(t)}\to\bHil_{\gamma(t)}.
	\end{equation}

	The mean value of $\dyn{A}$ at a state characterized by a state
vector $\psi$ or, equivalently, by  the corresponding to it
state lifting $\Psi_\gamma$ is
	\begin{equation}	\label{2.12}
\langle\ope{A}(t)\rangle_\psi^t
:= \frac{\langle\psi(t) | \ope{A}(t)\psi(t)\rangle}
	{\langle\psi(t) | \psi(t)\rangle}
= \left\langle \mor{A}_{\gamma}(t) \right\rangle_{\Psi_\gamma}^{t}
:=
\frac{
\langle \Psi_\gamma(t) | \mor{A}_\gamma(t) \Psi_\gamma(t) \rangle_{\gamma(t)}
}
{
\langle \Psi_\gamma(t) | \Psi_\gamma(t) \rangle_{\gamma(t)}
}.
	\end{equation}
Here
	\begin{equation}	\label{2.13}
\langle \cdot | \cdot \rangle_x =
\langle l_x \cdot | l_x \cdot \rangle, \qquad x\in\base
	\end{equation}
is the fibre Hermitian scalar product in $\HilB$ induced by  the Hermitian
scalar product
$\langle \cdot | \cdot \rangle \colon \Hil\times\Hil\to\mathbb{C}$
in $\Hil$.

	A summary of the above and other details concerning the Hilbert
space and Hilbert bundle description of (nonrelativistic) quantum mechanics
can be found in~\cite{bp-BQM-interpretation+discussion}.

\section {Time-dependent (Hamiltonian) approach}
\label{Sect3}

	Consider now the pure mathematical aspects of the scheme described in
Sect.~\ref{Sect2}.
	On one hand, it is essential to be notice that the Schr\"odinger
equation~\eref{2.1} is a first order (with respect to the time%
\footnote{%
The dependence of $\psi$ and $\Ham$ on the spatial coordinates (and momentum
operators) is inessential for the present section and, respectively, is not
written explicitly.%
}%
)
linear partial differential equation solved with respect to the time
derivative. On the other hand, the considerations of Sect.~\ref{Sect2} are
true for Hilbert spaces and bundles whose dimensionality is generically
infinity, but, as one can easily verify, they hold also for spaces and
bundles with finite dimension.

	These observations, as we shall prove below, are enough to transfer
the bundle nonrelativistic formalism to the relativistic region. We call the
result of this procedure
\emph{time\ndash dependent or Hamiltonian approach} as in
it the time plays a privileged r\^ole and the relativistic covariance is
implicit.

\subsection{General case}
\label{Subsect3.3}

	Taking into account the above, we can make the following conclusion.
Given a linear (vector) space $\fibre$ of $C^1$ functions
$\psi\colon J \to \mathbb{C}$ with $J$ being an
$\mathbb{R}$\ndash interval. (We do
not make any assumptions on the dimensionality of $\fibre$; it can be finite
as well as countable or uncountable infinity.) Let
$\Ham(t)\colon\fibre\to\fibre$, $t\in J$ be (possibly depending on $t$)
linear operator. Consider the equation%
\footnote{%
We introduce the multiplier $\ih\not=0$ from purely physical reasons and to be
able to apply the results already obtained directly, without any changes.%
}
	\begin{equation}	\label{3.1}
\ih\frac{\pd \psi(t)}{\pd t} = \Ham(t) \psi(t).
	\end{equation}
For it are valid all of the results of Sect.~\ref{Sect2}, viz., for instance,
there can be introduced the bundle $\bundle$, the evolution operator
$\ope{U}$, etc.; the relations between them being the same as in
Sect.~\ref{Sect2}. If the vector space $\fibre$ is endowed with a scalar
(inner) product
$\langle \cdot | \cdot \rangle \colon \fibre\times\fibre\to\mathbb{C}$,
then~\eref{2.13} induces analogous product in the bundle (\ie in the
bundle's fibres).
Consequently, imposing condition~\eref{2.12}, we get~\eref{2.11}. Further,
step by step, one can derive all of the results
of~\cite{bp-BQM-introduction+transport,bp-BQM-equations+observables,
	bp-BQM-pictures+integrals,bp-BQM-mixed_states+curvature,
	bp-BQM-interpretation+discussion,bp-BQM-full}.

	Further we shall need a slight, but very important generalization of
the above. Let now $\fibre$ be a vector space of vector-valued $C^1$ functions
$\psi\colon J \to V$ with $V$ being a complex vector space and
$\Ham(t)\colon\fibre\to\fibre$; the case just considered corresponds to
$V=\mathbb{C}$, i.e to $\dim_\mathbb{C}V=1$. It is not difficult to verify
that all of the above-said, corresponding to $\dim_\mathbb{C}V=1$, is also
\emph{mutatis mutandis} valid for $\dim_\mathbb{C}V\ge1$. Consequently, the
fibre bundle reformulation of the solution of~\eref{3.1}, the operators, and
scalar product(s) in $\fibre$ can be carried out in the general case when
$\psi\colon J\to V$ with $\dim_\mathbb{C}V\ge1$.

	Suppose now $\mathit{K}^{m}$ is the vector space of $C^m$,
$m\in\mathbb{N}\cup\{\infty\}\cup\{\omega\}$, vector-valued functions
$\varphi\colon J\to W$ with $W$ being a vector space. Let
$\varphi\in\mathit{K}^{m}$ satisfies the equation
	\begin{equation}	\label{3.2}
f\left(
t,\varphi,\frac{\pd\varphi}{\pd t},\ldots,\frac{\pd^n\varphi}{\pd t^n}
\right)
= 0
	\end{equation}
where
 \(
f\colon J\times
\mathit{K}^{m}\times\mathit{K}^{m-1}\times\cdots\times\mathit{K}^{m-n}
\to \mathit{K}^{m-n},
 \)
 $n\in\mathbb{N},\ n\le m$
is a map (multi)linear in $\varphi$ and its derivatives. We suppose~\eref{3.2}
to be solvable with respect to the highest derivative of $\varphi$,
i.e.~\eref{3.2} to be equivalent to%
\footnote{%
Requiring the equivalence of~\eref{3.2} and~\eref{3.3}, we exclude the
existence of sets on which the highest derivative of $\varphi$ entering
in~\eref{3.2} may be of order $k<n$. If we admit the existence of such sets,
then on them we have to replace $n$ in~\eref{3.3} with $k$. (Note $k$ may be
different for different such sets.) The below-described procedure can be
modified to include this more general situation but, since we do not want to
fill the presentation with complicated mathematical details, we are not going
to do this here.%
}
	\begin{equation}	\label{3.3}
\frac{\pd^n\varphi}{\pd t^n}
=
G\left(
t,\varphi,\frac{\pd\varphi}{\pd t},\ldots,\frac{\pd^{n-1}\varphi}{\pd t^{n-1}}
\right)
       	\end{equation}
for some map
\(
G\colon J\times
\mathit{K}^{m}\times\mathit{K}^{m-1}\times\cdots\times\mathit{K}^{m-n+1}
\to \mathit{K}^{m-n},
\)
linear in $\varphi$ and its derivatives.

	The already developed fibre bundle formalism for the
(Schr\"odinger-type) equation~\eref{3.1} can be transferred to~\eref{3.3}.
This can be done in a number of different ways. Below we shall realize the
most natural way, but one has to keep in mind that for a concrete equation
another method may turn to be more useful, especially from the view-point of
possible physical applications (see below Subsect.~\ref{Subsect3.5}).

	Defining
\(
\fibre :=
\mathit{K}^{m}\times\mathit{K}^{m-1}\times\cdots\times\mathit{K}^{m-n+1}
\)
and putting
	\begin{equation}	\label{3.3-1}
\psi :=
\left(
\varphi,\frac{\pd\varphi}{\pd t},\ldots,\frac{\pd^{n-1}\varphi}{\pd t^{n-1}}
\right)^\top
	\end{equation}
with $\top$ being the transposition sign, we can transform~\eref{3.3} in the
form~\eref{3.1} with `Hamiltonian'
	\begin{equation}	\label{3.4}
\Ham(t) = \ih \times
\begin{pmatrix}
0 & \id_{\mathit{K}^{m-1}} & 0                      & \ldots &  0 \\
0 & 0                      & \id_{\mathit{K}^{m-2}} & \ldots &  0 \\
\hdotsfor[2.11]{5} \\
0 & 0                      & 0                      & \ldots &
						\id_{\mathit{K}^{m-n+1}} \\
f_0(t)\id_{\mathit{K}^{m}} &
	f_1(t)\id_{\mathit{K}^{m-1}} &
		f_2(t)\id_{\mathit{K}^{m-2}} &
					\ldots &
					f_{n-1}(t)\id_{\mathit{K}^{m-n+1}}
\end{pmatrix}
	\ \colon\fibre\to\fibre
	\end{equation}
which is a linear matrix operator (see~\ref{AppendixB}), \ie a matrix of
linear operators. Here $\id_X$ is the identity map of a set $X$ and
$f_i\colon J\to\mathbb{C}$, $i=0,\ldots,n-1$ define the (multi)linear map
$G\colon J\times\fibre\to\mathit{K}^{m-n}$ by
	\begin{equation}	\label{3.5}
G\left(
t,\varphi,\frac{\pd\varphi}{\pd t},\ldots,\frac{\pd^{n-1}\varphi}{\pd t^{n-1}}
\right)
=
f_0(t)\varphi +
\sum_{i=1}^{n-1}f_i(t)\frac{\pd^i\varphi}{\pd t^i}.
	\end{equation}

	In this way we have proved that~\eref{3.2} can equivalently be
rewritten as a Schr\"odinger-type equation~\eref{3.1} with `Hamiltonian'
given by~\eref{3.4}. Such a transformation is not unique. For example, one
can choose the components of $\psi$ to be any $n$ linearly independent linear
combinations of
$\varphi$, $\pd\varphi/\pd t$, \ldots,  $\pd^{n-1}\varphi/\pd t^{n-1}$; this
will result only in another form of the matrix~\eref{3.4}.
In fact, if $A(t)$ is a nondegenerate matrix\ndash valued function, the
change
	\begin{equation}	\label{3.7}
\psi(t)\mapsto\widetilde{\psi}(t) = A(t)\psi(t)
	\end{equation}
leads to~\eref{3.1} with Hamiltonian
	\begin{equation}	\label{3.8}
\widetilde{\Ham}(t) = A(t)\Ham A^{-1}(t) +\frac{\pd A(t)}{\pd t}A^{-1}(t) .
	\end{equation}

	Now to the equation~\eref{3.1}, with $\Ham$ given via~\eref{3.4}, we
can apply the already described procedure for reformulation in terms of
bundles.

\subsection {Dirac equation}
\label{Subsect3.4}

	The relativistic quantum mechanics of spin $\frac{1}{2}$ particle is
described by the Dirac
equation (see~\cite[chapter~2]{Itzykson&Zuber},
\cite[chapters~1\nobreakdash--5]{Bjorken&Drell-1},
and~\cite[part~V, chapter~XX]{Messiah-2}).
The wave function $\psi=(\psi_1,\psi_2,\psi_3,\psi_4)^\top$ of such a
particle is a 4\ndash dimensional (4\ndash component) spinor satisfying this
equation whose Schr\"odin\-ger\ndash type form%
\footnote{%
For the purposes of this section we do not need the widely know relativistic
invariant form of Dirac
equation~\cite{Itzykson&Zuber, Bjorken&Drell-1, Messiah-2} (see also
Subsect.~\ref{Subsect4.7}).%
}
is (see, e.g.~\cite[chapter~XX, \S~6, equation~(36)]{Messiah-2}
or~\cite[chapter~1, \S~3, equation~(1.14)]{Bjorken&Drell-1})
	\begin{equation}	\label{4.1}
\ih\frac{\pd\psi}{\pd t} = \lindexrm[\Ham]{}{D} \psi
	\end{equation}
where $\lindexrm[\Ham]{}{D}$ is a Hermitian operator, called
\emph{Dirac Hamiltonian}, in the space $\fibre$ of state vectors (spinors).
For a spin $\frac{1}{2}$ particle with (proper) mass $m$ and electric charge
$e$ the explicit form of $\lindexrm[\Ham]{}{D}$ in an (external)
electromagnetic field with 4\ndash potential $(\varphi,\Vect{\mathit{A}})$
is~\cite[chapter~XX, \S~6, equation~(44)]{Messiah-2}
	\begin{equation}	\label{4.2}
\lindexrm[\Ham]{}{D} =
e\varphi\openone_4 +
c\Vect{\alpha} \cdot
	(\Vect{p} - \frac{e}{c} \Vect{\mathit{A}} ) + mc^2\beta,
	\end{equation}
where $\openone_4$ is the $4\times4$ unit matrix,
$c$ is the velocity of light in vacuum,
$\Vect{\alpha}:=(\alpha^1,\alpha^2,\alpha^3)$ is a matrix vector of
$4\times4$ matrices, $\Vect{p}=-\ih\boldsymbol{\nabla}$ is the
(3\ndash dimensional) momentum operator ($\nabla_i:=\pd/\pd x^i,\ i=1,2,3$),
and $\beta$ is a $4\times4$ matrix. The explicit (and general) forms of
 $\alpha^1,\alpha^2,\alpha^3$, and $\beta$ can be found, e.g.
in~\cite[chapter~2, sect.~2.1.2, equation~(2.10)]{Itzykson&Zuber}.

	Since~\eref{4.1} is a first-order equation, we can introduce the
\emph{Dirac evolution operator} $\lindexrm[\ope{U}]{}{D}$ via
 $\psi(t)=\lindexrm[\ope{U}]{}{D}(t,t_0)\psi(t_0)$ (cf.~\eref{2.2}).
Generally it is a $4\times4$ integral\ndash matrix operator
(see~\ref{AppendixB}) uniquely defined by the initial\ndash value problem
	\begin{equation}	\label{4.3}
\ih\frac{\partial}{\partial t} \lindexrm[\ope{U}]{}{D}(t,t_0) =
\lindexrm[\Ham]{}{D}(t) \circ \lindexrm[\ope{U}]{}{D}(t,t_0),
\qquad
\lindexrm[\ope{U}]{}{D}(t_0,t_0) = \id_\fibre
	\end{equation}
with $\fibre$ being the space of 4-spinors.
The explicit form of $\lindexrm[\ope{U}]{}{D}$ is derived, e.g.\
in~\cite[chapter~2, sect.~2.5.1]{Itzykson&Zuber}.

	Now the bundle formalism developed
in~\cite{bp-BQM-introduction+transport,bp-BQM-equations+observables,
	bp-BQM-pictures+integrals,bp-BQM-mixed_states+curvature,
	bp-BQM-interpretation+discussion,bp-BQM-full}
can be applied to a description of Dirac particles practically without
changes. For instance, the spinor lifting of paths have to be introduced
via~\eref{2.3} and are connected by~\eref{2.5} in which $\mor{U}$ has to be
replaced by the \emph{Dirac evolution transport}
$\lindexrm[\mspace{-2.2mu}\mor{U}]{}{D}$ given by
	\begin{equation*}
\lindexrm[\mspace{-2.2mu}\mor{U}]{}{D}_\gamma(t,s) =
   l_{\gamma(t)}^{-1}\circ \lindexrm[\ope{U}]{}{D}(t,s) \circ l_{\gamma(s)},
\qquad s,t\in J
	\end{equation*}
(cf.~\eref{2.4}). The \emph{bundle Dirac equation} is
	\begin{equation}	\label{4.4}
\lindexrm[\mspace{-2.2mu}\mor{D}]{}{D}_{t}^{\gamma} \Psi_\gamma = 0
	\end{equation}
with $\lindexrm[\mspace{-2.2mu}\mor{D}]{}{D}$ being the assigned to
$\lindexrm[\mspace{-2.2mu}\mor{U}]{}{D}$
by~\eref{2.7} derivation along paths. Again, the matrix of the coefficients
of the Dirac evolution transport is connected with the matrix-bundle Dirac
Hamiltonian via~\eref{2.10}, etc.

\subsection {Klein-Gordon equation}
\label{Subsect3.5}

	The wavefunction $\phi\in\mathit{K}^m$ of spinless special\ndash
relativistic particle is a scalar function of class $C^m,\ m\geq2$, over the
spacetime and satisfies the Klein-Gordon
equation~\cite[chapte~9]{Bjorken&Drell-1}.  For a particle of mass $m$ and
electric charge $e$ in an external electromagnetic field with 4\ndash
potential $(\varphi,\Vect{\mathit{A}})$ it
reads~\cite[chapter~XX, \S~5, equation~(30)]{Messiah-2}
	\begin{equation}	\label{5.1}
\Bigl[
\Bigl(\ih\frac{\pd}{\pd t} -e\varphi\Bigr)^2
-
c^2 \left( \Vect{p} - \frac{e}{c}\Vect{\mathit{A}} \right)^2
\Bigr]
\phi
=m^2c^4\phi.
	\end{equation}

	This is a second\ndash order linear partial differential equation of
type~\eref{3.2} with respect to $\phi$. Solving it with respect to
$\pd^2\phi/\pd t^2$, we can transform it to equation of type~\eref{3.3}:
	\begin{equation}	\label{5.2}
	\begin{split}
&	\frac{\pd^2\phi}{\pd t^2}
=
f_0\phi + \frac{2e}{\ih}\varphi\frac{\pd\phi}{\pd t} \\
&	f_0 :=
\left[
-\frac{c^2}{\hbar^2}
\left( \Vect{p} - \frac{e}{c}\Vect{\mathit{A}} \right)^2
-
\frac{m^2c^4}{\hbar^2}
+
\frac{e^2}{\hbar^2}\varphi^2
+
\frac{2e}{\ih}\frac{\pd\varphi}{\pd t}
\right]
\id_{\mathit{K}^m}.
	\end{split}
	\end{equation}

	As pointed above, there are (infinitely many) different ways to put
this equation into Schr\"odinger\nobreakdash-type form. Below we realize
three of them, each having applications for different purposes.

	The `canonical' way to do this is to define
 $\psi := (\phi,\pd\phi/\pd t)^\top$ and
 \(
\mathit{K}^m :=
	\{ \phi\colon J\to\mathbb{C},\quad \phi\text{ is of class } C^m \},\
	m\geq2.
 \)
Then, comparing~\eref{5.2} with~\eref{3.5}, we see that~\eref{5.2}, and
hence~\eref{5.1}, is equivalent to~\eref{3.1} with
$\Ham=\lindexrm[\Ham]{\mspace{32mu}c}{K-G}$, where the `canonical'
Klein\ndash Gordon Hamiltonian is

	\begin{equation}	\label{5.3}
\lindexrm[\Ham]{\mspace{32mu}c}{K-G}
:=
\ih
	\begin{pmatrix}
0   & \id_{\mathit{K}^{m}}			 \\
f_0 & \frac{2e}{\ih}\varphi\id_{\mathit{K}^{m-1}}
	\end{pmatrix}
~~~.
	\end{equation}
It should be note, for a free scalar particle,
$(\varphi,\Vect{\mathit{A}})=0$, this is the anti\ndash diagonal matrix
operator
\(
\left(\begin{smallmatrix}
0                            & 1  \\
c^2\nabla^2 - m^2c^4/\hbar^2 & 0
\end{smallmatrix}\right)
\id_{\mathit{K}^{m}}
\).

	Another possibility is to put
\(
\psi=\bigl( \phi + \frac{\ih}{mc^2}\frac{\pd\phi}{\pd t},\phi -
	\frac{\ih}{mc^2}\frac{\pd\phi}{\pd t} \bigr)^\top.
\)
This choice is good for investigation of the non-relativistic limit, when
$\ih\frac{\pd\phi}{\pd t}\approx mc^2\phi$~\cite[chapterr~XX,
\S~5]{Messiah-2}, so that in it $\psi\approx(2\phi,0)^\top$.

	Now, as a simple verification proves, the Schr\"odinger\ndash type
form~\eref{3.1} of~\eref{5.2} is realized for the Hamiltonian
	\begin{equation*}
\lindexrm[\Ham]{\mspace{13.0mu}n.r.}{K-G} := \frac{1}{2} \times
	\begin{pmatrix}
(mc^2 + 2e\varphi)\id_{\mathit{K}^{m-1}} - \hbar^2f_0/mc^2 &
	(- mc^2 - 2e\varphi)\id_{\mathit{K}^{m-1}} - \hbar^2f_0/mc^2 \\
(mc^2 - 2e\varphi)\id_{\mathit{K}^{m-1}} + \hbar^2f_0/mc^2 &
	(- mc^2 + 2e\varphi)\id_{\mathit{K}^{m-1}} + \hbar^2f_0/mc^2
	\end{pmatrix}~.
	\end{equation*}

	If the electromagnetic field vanishes,
 $(\varphi,\Vect{\mathit{A}})=0$, then
 $f_0=(c^2\nabla^2-m^2c^4/\hbar^2)\id_{\mathit{K}^{m}}$ and
	\begin{multline*}
\lindexrm[\Ham]{\mspace{13mu}n.r.}{K-G} = \frac{1}{2}
\left[
mc^2
	\begin{pmatrix}
1 &  0 \\
0 & -1
	\end{pmatrix}
- \frac{\hbar^2}{2m}\boldsymbol{\nabla}^2
	\begin{pmatrix}
1  &  1 \\
-1 & -1
	\end{pmatrix}
\right] \id_{\mathit{K}^{m-1}}
					\\
=
\left[
\Bigl( mc^2 + \frac{\Vect{p}^2}{2m} \Bigr) \mspace{3mu}
	\begin{pmatrix}
1  &  0 \\
0  & -1
	\end{pmatrix}
+ \frac{\Vect{p}^2}{2m}
	\begin{pmatrix}
0  &  1 \\
-1 & 0
	\end{pmatrix}
\right] \id_{\mathit{K}^{m-1}}
	\end{multline*}

	The third possibility mentioned corresponds to the choice of $\psi$
as a $5\times1$ matrix:
\[
\psi =
\bigl(
mc^2\phi, \frac{\pd\phi}{\pd t}, \frac{\pd\phi}{\pd x^1},
\frac{\pd\phi}{\pd x^2}, \frac{\pd\phi}{\pd x^3}
\bigr)^\top
=
\bigl( mc^2\phi, \frac{\pd\phi}{\pd t}, \boldsymbol{\nabla}\phi \bigr)^\top.
\]
The corresponding Hamiltonian is $5\times5$ matrix which in the absence of
electromagnetic field is~\cite[chapter~2, sect.~2.1.1]{Itzykson&Zuber}
\(
\lindexrm[\Ham]{\mspace{30.2mu}5}{K-G}
= mc^2\beta+\Vect{\alpha}\cdot\boldsymbol{\nabla}
\)
(cf.\ Dirac Hamiltonian~\eref{4.2} for $(\varphi,\Vect{\mathit{A}})=0$)
where $\beta$ is $5\times5$ matrix and $\Vect{\alpha}$ is
a 3\ndash vector of  $5\times5$ matrices. The full realization of this
procedure and the explicit form of the corresponding $5\times5$ matrices is
given in~\cite[\S~4, sect.~4.4]{Bogolyubov&Shirkov} (see also below
Subsect.~\ref{Subsect4.8}, formulae~\eref{8.2}).

	Choosing some representation of Klein\ndash Gordon equation as
first\ndash order (Schr\"odinger\nobreakdash-type) equation, we can in an
evident way transfer the bundle formalism to the description of spinless
particles.

\subsection {Other relativistic wave equations}
\label{Subsect3.6}

	As we saw in subsections~\ref{Subsect3.4} and~\ref{Subsect3.5}, the
only problem for a bundle reformulation of a wave equation is to rewrite it
as a first\nobreakdash-order differential equation and to find the
corresponding Hamiltonian.  Since all relativistic wave equations are of the
form of
equation~\eref{3.2}~\cite{Bogolyubov&Shirkov,Messiah-2,Bjorken&Drell-1},
this procedure can successfully be performed for all of them.

	For instance, the wavefunction
$\psi=(\psi_0,\psi_1,\psi_2,\psi_3)^\top$ of particles with spin~1 is a
4\ndash vector satisfying Klein-Gordon equation%
\footnote{%
Here we do not concern the additional conditions like the Lorentz one. They
lead to a modification of the equations defining the evolution transport, but
this does not change the main ideas. E.g.\ the most general equation for
vectorial mesons is the Proca
equation~\cite[chapter~3, equation~(3.132)]{Itzykson&Zuber}
\(
\bigl[
\bigl( \partial^\varkappa\partial_\varkappa + \frac{m^2c^4}{\hbar^2} \bigr)
\eta_{\mu\nu}
-
\partial_\mu\partial_\nu
\bigr]
\phi^\nu
= 0
\)
on which one usually imposes the additional condition $\pd_\nu\phi^\nu=0$.
(For $m\not=0$ the last condition is a corollary of the Proca equation; to
prove this, simply apply $\pd^\mu$ to Proca equation.)%
}%
~\cite[capter~I, \S~4]{Bogolyubov&Shirkov}. Hence for each component
$\psi_i$, $i=0,1,2,3$ we can construct the corresponding Hamiltonian
$\lindexrm[\Ham]{}{K-G}_i$ using, e.g., one of the methods described in
Subsect.~\ref{Subsect3.5}.
Then equation~\eref{3.1} holds for a Hamiltonian of the form of a
$4\times4$ diagonal block matrix operator
\[
\lindexrm[\Ham]{}{K-G}
	= \diag\bigl( \lindexrm[\Ham]{}{K-G}_0, \lindexrm[\Ham]{}{K-G}_1,
	\lindexrm[\Ham]{}{K-G}_2, \lindexrm[\Ham]{}{K-G}_3 \bigr)
\]
and the corresponding new wavefunction which is now a $8\times1$ matrix.

	The just said is, of course, valid with respect to electromagnetic
field, the 4\ndash potential playing the r\^ole of
wavefunction~\cite[chapter~I, \S~5]{Bogolyubov&Shirkov}. It is interesting to
be noted that even at a level of classical electrodynamics the Maxwell
equations admit Schr\"odinger\nobreakdash-like form. There are different
fashions to do this.
For a free field, one of them is to put
\(
\psi = (\Vect{E},\Vect{H},\Vect{E},\Vect{H})^\top
\)
where $\Vect{E}$ and $\Vect{H}$ are respectively the electric and
magnetic field strengths, and to define the Hamiltonian, e.g., by
\(
\lindexrm[\Ham]{}{E-M} = \ih
	\left(	\begin{smallmatrix}
0	& c\rot	& 0	& 0	\\
-c\rot	& 0	& 0	& 0	\\
0	& 0	& 0	& \diver\\
0	& 0	&-\diver& 0
	\end{smallmatrix}	\right)~.
\)

	Almost the same as spin~1 particles is the case of particles with
spin~2. Their wavefunction is symmetric tensor field of second rank whose
components satisfy also the Klein\ndash Gordon equation~\cite[p.~24]{Nelipa}.

	The wavefunction of particles with spin $3/2$ is a
(4\nobreakdash-)spin\ndash vector whose `vector' components satisfy the Dirac
equation (and some additional conditions)~\cite[pp.~35\Ndash 36]{Nelipa}.
Therefore for them can be applied \emph{mutatis mutandis} the presented in
Subsect.~\ref{Subsect3.4} bundle formalism.

\section {Covariant approach}
\label{Sect4}

	The developed in Sect.~\ref{Sect3} bundle formalism for relativistic
quantum wave equations has the deficiency that it is not explicitly covariant;
so it is not in harmony with the relativistic theory it represents. This is
a consequence of the direct applications of the bundle methods developed for
the nonrelativistic region, where they work well enough, to the relativistic
one. The present section is intended to mend this `defect'. Here we develop
an appropriate covariant bundle description of relativistic quantum mechanics
which corresponds to the natural character of this theory. The difference
between the time\ndash dependent and covariant approach is approximately the
same as the one between the Hamiltonian and Lagrangian derivation (and forms)
of some relativistic wave equation.

\subsection {Dirac equation}
\label{Subsect4.7}

	The covariant Dirac
equation~\cite[sect.~2.1.2]{Itzykson&Zuber},
\cite[chapter~XX, \S~8]{Messiah-2}
for a (spin $\frac{1}{2}$) particle with mass $m$ and electric charge $e$ in
an external electromagnetic field with 4\ndash potential $\ope{A}^\mu$ is
	\begin{equation}	\label{7.1}
(\ih\Slashed{D} - mc\openone_4) \psi = 0,
\qquad
\Slashed{D}:=\gamma^\mu D_\mu, \quad
			D_\mu:=\pd_\mu -\frac{e}{\ih c}\ope{A}_\mu .
	\end{equation}
Here $i\in\mathbb{C}$ is the imaginary unit, $\hbar$ is the Planck constant
(divided by $2\pi$),
$\openone_4=\diag(1,1,1,1)$ is the $4\times4$ unit matrix,
$\psi:=(\psi^0,\psi^1,\psi^2,\psi^3)$ is (the matrix of the components of) a
4\ndash spinor, $\gamma^\mu$, $\mu=0,1,2,3$, are the well known Dirac
$\gamma$\ndash matrices~\cite{Bjorken&Drell-1, Messiah-2,Itzykson&Zuber}, and
$c$ is the velocity of light in vacuum.
	Since~\eref{7.1} is a first order partial differential equation on the
Minkowski 4\ndash dimensional spacetime, it does not admit an evolution
operator with respect to the spacetime. More precisely, if $x_1,x_2\in M$,
$M$ being the spacetime, \ie the Minkowski space $M^4$, then there does
\emph{not} exist a $4\times4$ matrix operator (see~\ref{AppendixB})
$\ope{U}(x_2,x_1)$ such that
	\begin{gather*}	
\psi(x_2) = \ope{U}(x_2,x_1) \psi(x_1).
\\
(\ih\Slashed{D}_x - mc\openone_4) \ope{U}(x,x_0) = 0, \qquad
\ope{U}(x_0,x_0) = \id_\fibre, \quad x,x_0\in M
	\end{gather*}
where
$\Slashed{D}=\slashed{\pd} - \frac{e}{\ih c} \Slashed{\ope{A}}$
with
$\slashed{\pd}:=\gamma^\mu\pd/\pd x^\mu$ and
$\Slashed{\ope{A}}:=\gamma^\mu\ope{A}_\mu$,
$\fibre$ is the space of 4\ndash spinors,
and $\id_X$ is the identity map of a set $X$.

	By this reason, the methods of previous sections cannot
be applied directly to the general 4-dimensional spacetime descriptuion of
the Dirac equation. An altternative approach to the problem is presented
below.

	Suppose $\bundle$ is a vector bundle with (total) bundle space
$\bspace$, projection $\pr\colon\bspace\to\bbase$, fibre $\fibre$,
and isomorphic fibres $\fibreover{x}:=\pr^{-1}(x)$, $x\in\bbase$.
There exist linear isomorphisms $l_x:\fibreover{x}\to\fibre$ which
we assume to be diffeomorphisms; so
$\fibreover{x}=l_{x}^{-1}(\fibre)$ are 4\ndash dimensional vector spaces.

	To a state vector (spinor) $\psi(x_0)$ at a fixed point $x_0$, we
assign a $C^1$ section
\footnote{%
In contrast to the time\ndash dependent approach (Sect.~\ref{Sect3}) and
nonrelativistic case (Sect.~\ref{Sect2}) now $\Psi_{x_0}$ is simply a
section, not section along paths~\cite{bp-BQM-introduction+transport}.
Physically this corresponds to the fact that quantum objects do not have
world lines (trajectories) in a classical sense~\cite{Messiah-1}.%
}
$\Psi_{x_0}$ of $\bundle$, i.e.\ $\Psi_{x_0}\in\Sec^1\bundle$, by
(cf.~\eref{2.3})
	\begin{equation}	\label{7.4}
\Psi_{x_0}(x)
:= l_{x}^{-1}\bigl(\psi(x_0)\bigr) \in\fibreover{x}:=\pr^{-1}(x),
\quad x\in M.
	\end{equation}
Generally $\Psi_{x_0}(x)$ depends on the choice of the point $x_0\in\bbase$.

	Since in $\bundle$ the state of a Dirac particle is described by
$\Psi_{x_0}$, we call it \emph{state section}; resp.\ $\bundle$ is the
\emph{4\ndash spinor bundle}.
The description of Dirac particle
via $\Psi_{x_0}$ will be called \emph{bundle description}. If it is known, the
conventional spinor description is achieved by the spinor
	\begin{equation}	\label{7.4'}
\psi(x_0) := l_{x}\bigl(\Psi_{x_0}(x)\bigr) \in\fibre.
	\end{equation}
Evidently, we have
	\begin{equation}	\label{7.5}
\Psi_{x_0}(x_2) = \mor{L}(x_2,x_1) \Psi_{x_0}(x_1),\qquad x_2,x_1\in M.
	\end{equation}
with (cf.~\eref{2.4})
	\begin{equation}	\label{7.6}
\mor{L}(y,x) :=
l_{y}^{-1} \circ l_{x}\colon\fibreover{x}\to\fibreover{y},
\qquad x,y\in M.
	\end{equation}

	Obviously, $\ope{L}$ is a linear $4\times4$ matrix operator
satisfying the equations
	\begin{align}	\label{7.8}
\mor{L}(x_3,x_1) &= \mor{L}(x_3,x_2) \circ \mor{L}(x_2,x_1),
\qquad x_1,x_2,x_3\in M,
\\
	\label{7.9}
\mor{L}(x,x) &= \id_{\fibreover{x}}, \qquad x\in M.
	\end{align}

	Consequently, by definition~\ref{DefnLT.1}, the map
$\mor{L}\colon(y,x)\to\mor{L}(y,x)=\mor{K}_{x\to y}^{\id_M}$ is a linear
transport along the identity map $\id_M$ of $M$ in the bundle $\bundle$.
Alternatively, as it is mentioned in~\ref{AppendixA}, this means that
\(
\mor{L}^{\gamma}\colon(t,s) \to
\mor{L}_{s\to t}^{\gamma}:=\mor{L}(\gamma(t),\gamma(s)),\
s,t\in J
\)
is a flat linear transport along $\gamma\colon J\to M$ in $\bundle$.
(Besides, $\mor{L}(y,x)$ is a Hermitian and unitary transport ---
see~\cite{bp-BQM-introduction+transport}.)

	Equation~\eref{7.5} simply means that $\Psi_{x_0}$ is
$\mor{L}$\ndash\emph{transported (along $\id_M$) section} of $\bundle$
(cf.~\cite[definition~5.1]{bp-LTP-appl}). Writing~\eref{LT.10} for the
transport $\mor{L}$ and applying the result to a state section
given by~\eref{7.4}, one can prove that~\eref{7.5} is equivalent to
	\begin{equation}	\label{7.10}
\mathcal{D}_\mu \Psi = 0, \qquad \mu=0,1,2,3
	\end{equation}
where, for brevity, we have put
$\mathcal{D}_\mu:=\mathcal{D}_{x^\mu}^{\id_M}$ which is the
$\mu$\nobreakdash-th partial (section\ndash)derivation along the identity map
(of the spacetime) assigned to the transport $\mor{L}$.

	Now we shall introduce local bases and take a local view of the
above\ndash described material.

	Let $\{f_\mu(x)\}$ be a basis in $\fibre$ and $\{e_\mu(x)\}$ be a
basis in $\fibreover{x},\ x\in M$.
The matrices corresponding to vectors and/or
linear maps (operators) in these fields of bases will be denoted by the same
(kernel) symbol but in \textbf{boldface}, for instance:
$\Mat{\psi}:=\bigl(\psi^0,\psi^1,\psi^2,\psi^3\bigr)^\top$ and
$\Mat{l}_{x}(y):=\bigl[(l_x)_{~\nu}^{\mu}(y)\bigr]$
are defined, respectively, by
 $\psi(x)=:\psi^\mu(x)f_\mu(x)$ and
$l_x\bigl(e_\nu(x)\bigr)=:\bigl(l_x(y)\bigr)_{~\nu}^{\mu}f_\mu(y)$.
We put $\Mat{l}_x:=\Mat{l}_x(x)$; in fact this will be the only case when the
matrix of $l_x$ will be required as we want the `physics in
$\fibreover{x}$' to
correspond to that of $\fibre$ at $x\in M$. A very convenient choice is to
put $e_\mu(x)=l_{x}^{-1}\bigl(f_\mu(x)\bigr)$; so then
$\Mat{l}_x=\openone_4:=\bigl[\delta_{\nu}^{\mu}\bigr]=\diag(1,1,1,1)$ is the
$4\times4$ unit matrix.

	The matrix elements of the mapping $\mor{L}(y,x)$ are defined via the
equation
$\mor{L}(y,x)\bigl(e_\mu(x)\bigr)=:\mor{L}_{~\mu}^{\lambda}(y,x)e_\lambda(y)$
and, due to~\eref{7.6}, we have
	\begin{equation}	\label{7.12}
\Mat{\mor{L}}(y,x) =
	\Mat{l}_{y}^{-1} \cdot \Mat{l}_{x}
	\end{equation}
which is generically a matrix operator (see~\ref{AppendixB}).

	According to~\eref{LT.12} the \emph{coefficients} of the
transport $\mor{L}$ form four matrix operators
	\begin{equation}	\label{7.13}
\lindex[\Mat{\Gamma}]{\mu}{}(x) :=
\bigl[ \lindex[\Gamma]{\mu}{} _{~\nu}^{\lambda}(x) \bigr]_{\lambda,\nu=0}^{3}
  := \left. \frac{\pd\Mat{\mor{L}}(x,y)}{\pd y^\mu}\right|_{y=x}
   = \Mat{l}^{-1}(x) \frac{\pd\Mat{l}(x)}{\pd x^\mu}
	\end{equation}
where~\eref{7.12} was applied (see also theorem~\ref{ThmLT.1}).

	Applying~\eref{LT.5} and~\eref{7.13}, we find
	\begin{equation}	\label{7.14}
\frac{\pd\Mat{\mor{L}}(y,x)}{\pd y^\mu} =
- \lindex[\Mat{\Gamma}]{\mu}{}(y) \odot \Mat{\mor{L}}(y,x), \qquad
\frac{\pd\Mat{\mor{L}}(y,x)}{\pd x^\mu} =
\Mat{\mor{L}}(y,x) \odot \lindex[\Mat{\Gamma}]{\mu}{}(x) ,
	\end{equation}
where $\odot$ denotes the introduced by~\eref{mo.2} multiplication of matrix
operators. Therefore
	\begin{equation}	\label{7.15}
\slashed{\pd}_y \Mat{\mor{L}}(y,x)  =
	- \Slashed{\Mat{\Gamma}}(y) \Mat{\mor{L}}(y,x),
\qquad
\Slashed{\Mat{\Gamma}}(x) :=
	\gamma^\mu \cdot \lindex[\Mat{\Gamma}]{\mu}{}(x) .
	\end{equation}

	Similarly to the nonrelativistic
case~\cite{bp-BQM-equations+observables}, to any operator
$\ope{A}\colon\fibre\to\fibre$ we assign a bundle morphism
$\mor{A}\colon\bspace\to\bspace$ by
	\begin{equation}	\label{7.16}
\mor{A}_x := \mor{A}|_{\fibreover{x}} := l_{x}^{-1}\circ\ope{A}\circ l_x.
	\end{equation}
Defining
	\begin{equation}	\label{7.17}
\mor{G}^\mu(x) := l_{x}^{-1}\circ\gamma^\mu\circ l_x,
\quad
d_\mu := d_\mu|_x :=  l_{x}^{-1}\circ\pd_\mu\circ l_x,
\qquad
\pd_\mu:=\frac{\pd}{\pd x^\mu}
	\end{equation}
and using the matrices $\ope{G}^\mu(x)$ and $\ope{E}_\mu(x)$ given
via~\eref{mo.11}, we get
	\begin{equation}	\label{7.18}
\Mat{\mor{G}}^\mu(x) = \Mat{l}_{x}^{-1}\ope{G}^\mu(x) \Mat{l}_x,
\quad
\Mat{d}_\mu
:= \openone_4\pd_\mu +
   \Mat{l}_{x}^{-1}\bigl(\pd_\mu\Mat{l}_x + \ope{E}_\mu(x)\Mat{l}_x \bigr).
	\end{equation}
(Here and below, for the sake of shortness, we sometimes omit the argument
$x$.)
	The anticommutation relations
	\begin{equation}	\label{7.19}
	\begin{split}
\mor{G}^\mu\mor{G}^\nu + \mor{G}^\nu\mor{G}^\mu
& = 2\eta^{\mu\nu}\id_\bspace,
\\ 
\Mat{\mor{G}}^\mu\Mat{\mor{G}}^\nu + \Mat{\mor{G}}^\nu\Mat{\mor{G}}^\mu
& = \ope{G}^\mu\ope{G}^\nu + \ope{G}^\nu\ope{G}^\mu
= 2\eta^{\mu\nu}\openone_4
\quad
\bigl(= \gamma^\mu\gamma^\nu + \gamma^\nu\gamma^\mu \bigr) ,
	\end{split}
	\end{equation}
where $[\eta^{\mu\nu}]=\diag(1,-1,-1,-1)=[\eta_{\mu\nu}]$ is the Minkowski
metric tensor, can be verified by means of~\eref{7.17}, \eref{7.18},
\eref{mo.9}, and the well know analogous relation for the
$\gamma$\ndash matrices (see, e.g.~\cite[chapter~2,
equation~(2.5)]{Bjorken&Drell-1}).

	For brevity, if $a_\mu\colon\bspace\to\bspace$ are morphisms, sums
like $\mor{G}^\mu\circ a_\mu$ will be denoted by `backslashing' the kernel
letter, $\backslashed{a}:=\mor{G}^\mu\circ a_\mu$ (\cf the `slashed' notation
$\slashed{a}:=\gamma^\mu a_\mu$). Similarly, we put
$\backslashed{\Mat{a}}:=\Mat{\mor{G}}^\mu(x)\Mat{a}_\mu$ for
$\Mat{a}_\mu\in GL(4,\mathbb{C})$.
It is almost evident (see~\eref{7.17}) that the morphism corresponding to
$\slashed{\pd}:=\gamma^\mu\pd_\mu$ is
$\backslashed{d}=\mor{G}^\mu(x)\circ d_\mu$:
	\begin{equation}	\label{7.20}
\backslashed{d} = l_{x}^{-1} \circ \slashed{\pd} \circ l_x .
	\end{equation}

	Now we can write the Dirac equation~\eref{7.1} in a bundle form.

	First of all, we rewrite~\eref{7.1} as
	\begin{equation}	\label{7.22}
\ih\slashed{\pd}_x\psi(x) = \mathcal{D}_x\psi(x),
\qquad
\mathcal{D}_x := mc\openone_4 + \frac{e}{c}\Slashed{\ope{A}}(x) ,
	\end{equation}
the index $x$ meaning that the corresponding operators act with respect to
the variable $x\in\bbase$. This is the \emph{covariant}
Schr\"odinger\nobreakdash-like form of Dirac equation; $\slashed{\pd}_x$ is
the analogue of the time derivation $\od/\od t$ and $\mathcal{D}_x$
corresponds to the Hamiltonian $\Ham$. We call $\mathcal{D}$ the \emph{Dirac
function}, or simply, \emph{Diracian} of a particle described by Dirac
equation.

	Substituting~\eref{7.4'} into~\eref{7.22}, acting on the result from
the left by $l_{y}^{-1}$, and using~\eref{7.17}, we find the bundle form
of~\eref{7.22} as
	\begin{equation}	\label{7.23}
\ih\backslashed{d}_x|_y \Psi_x(y) = D_x|_y\Psi_x(y)
	\end{equation}
with $y\in\bbase$, $\backslashed{d}_x|_y:=\mor{G}(y) {d_\mu}|_x|_y$,
 ${d_\mu}|_x|_y:=l_y^{-1}\circ\frac{\pd}{\pd x^\mu}\circ l_y$
$D_x\in\Morf\bundle$ being the (Dirac) bundle morphism assigned to the
Diracian. We call it the \emph{bundle Diracian}. According to~\eref{7.16} it
is defined by
	\begin{equation}	\label{7.24}
D_x|_y := D_x|_{\fibreover{y}} = l_{y}^{-1}\circ\mathcal{D}_x\circ l_y
     = mc\id_{\fibreover{y}} + \frac{e}{c}\Backslashed{\mor{A}}_x|_y ,
	\end{equation}
where
 $\mor{A}_\mu|_x=\ope{A}_\mu|_x \id_\bspace\in\Morf\bundle$,
\(
\mor{A}_\mu|_x|_y
=
l_{y}^{-1}\circ(\ope{A}_\mu|_x\id_\fibre)\circ l_y
=
\ope{A}_\mu|_x\id_{\fibreover{y}}
\)
are the components of the \emph{bundle electromagnetic potential} and
	\begin{multline}	\label{7.25}
\Backslashed{\mor{A}}_x |_y
= \mor{G}^\mu(y)\circ\mor{A}_\mu|_x |_y
= \mor{G}^\mu(y)\circ (l_y^{-1}\circ \ope{A}_\mu\id_\fibre \circ l_y )
= l_{y}^{-1}\circ( \Slashed{\ope{A}}|_x\id_\fibre )\circ l_y
= \Backslashed{\ope{A}}_x |_y.
	\end{multline}

\subsection {Other relativistic wave equations}
\label{Subsect4.8}

	The relativistic-covariant Klein-Gordon equation for a (spinless)
particle of mass $m$ and electric charge $e$ in a presence of (external)
electromagnetic field with 4\ndash potential $\ope{A}_\mu$
is~\cite[chapter~XX, \S~5, equation~(30$^\prime$)]{Messiah-2}
	\begin{equation}	\label{8.1}
\Bigl( \mathcal{D}^\mu\mathcal{D}_\mu +\frac{m^2c^2}{\hbar^2} \Bigr)\phi = 0,
\qquad
\mathcal{D}_\mu = \eta_{\mu\nu}\mathcal{D}^\nu
:=\pd_\mu - \frac{e}{\ih c}\ope{A}_\mu.
	\end{equation}
Since this is a second-order partial differential equation, it does not
directly admit an evolution operator and adequate bundle formulation and
interpretation. To obtain such a formulation, we have to rewrite~\eref{8.1}
as a first\ndash order (system of) partial differential equation(s) (\cf
Subsect.~\ref{Subsect3.5}).

	Perhaps the best way to do this is to replace $\phi$ with a
$5\times1$ matrix $\varphi=(\varphi^0,\ldots,\varphi^4)^\top$ and to
introduce $5\times5$ $\Gamma$\ndash matrices $\Gamma^\mu$,  $\mu=0,1,2,3$
with components $\bigl(\Gamma^\mu\bigr)_{~j}^{i}$, $i,j=0,1,2,3,4$ such that
(cf.~\cite[chapter~I, equations~(4.38) and~(4.37)]{Bogolyubov&Shirkov}):
	\begin{equation}	\label{8.2}
\varphi =
a
	\begin{pmatrix}
\ih\mathcal{D}_0\phi \\ \ih\mathcal{D}_1\phi \\
\ih\mathcal{D}_2\phi \\ \ih\mathcal{D}_3\phi \\ mc\phi
	\end{pmatrix},
\quad
\bigl(\Gamma^\mu\bigr)_{~j}^{i}
=	\begin{cases}
	1& 		\text{for $(i,j)=(\mu,4)$}	\\
	\eta_{\mu\mu}&	\text{for }(i,j)=(4,\mu)	\\
	0&		\text{otherwise}
	\end{cases}
	\end{equation}
where the complex constant $a\not=0$ is insignificant for us and can be
(partially) fixed by an appropriate normalization of $\varphi$.

	Then a simple checking shows that~\eref{8.1} is equivalent to
(cf.~\eref{7.1})
	\begin{equation}	\label{8.3}
(\ih\Gamma^\mu\mathcal{D}_\mu - mc\openone_5) \varphi = 0
	\end{equation}
with $\openone_5=\diag(1,1,1,1,1)$ being the $5\times5$ unit matrix, or
(cf.~\eref{7.22})
	\begin{equation}	\label{8.4}
\ih\Gamma^\mu\pd_\mu\varphi = \ope{K}\varphi,
\qquad
\ope{K} := mc\openone_5 + \frac{e}{c} \, \Gamma^\mu\ope{A}_\mu.
	\end{equation}

	Now it is evident that \emph{mutatis mutandis}, taking $\Gamma^\mu$
for $\gamma^\mu$, $\varphi$ for $\psi$, etc., the (bundle) machinery
developed in Subsect.~\ref{Subsect4.7} for the Dirac equation can be applied
to the Klein\ndash Gordon equation in the form~\eref{8.3}. Since the
transferring of the results obtained in Subsect.~\ref{Subsect4.7} for Dirac
equation to Klein\ndash Gordon one is absolutely trivial,%
\footnote{%
Both coincide up to notation or a meaning of the corresponding symbols.%
}
we are not going to present  here the bundle description of the latter
equation.

	Since the relativistic wave equations for particles with spin greater
than $1/2$ are versions or combinations of Dirac and Klein\ndash Gordon
equations~\cite{Nelipa,Bjorken&Drell-1,Messiah-2,Bogolyubov&Shirkov}, for them
is \emph{mutatis mutandis} applicable the bundle approach developed in
Subsect.~\ref{Subsect4.7} for Dirac equation or/and its version for
Klein\ndash Gordon  one pointed above.

\section
[Propagators and evolution transports or operators]
{Propagators and \\ evolution transports or operators}
\label{Sect9}

	The propagators, called also propagator functions or Green functions,
are solutions of the wave equations with point\nobreakdash-like unit source
and satisfy appropriate (homogeneous) boundary conditions corresponding to
a concrete problem under exploration~\cite{Bjorken&Drell-1,Itzykson&Zuber}.
Undoubtedly these functions play an important r\^ole in the mathematical
apparatus of (relativistic) quantum mechanics and its physical
interpretation~\cite{Bjorken&Drell-1,Itzykson&Zuber}. By this reason it is
essential to be investigated the connection between propagators and evolution
operators or/and transports. As we shall see below, the latter can be
represented as integral operators whose kernel is connected in a simple way
with the corresponding propagator.

\subsection{Green functions (review)}
\label{Subsect9.0}

	Generally~\cite[article ``Green function'']{Physicedia-1} the
\emph{Green function} $g(x',x)$ of a linear differential  operator $L$ (or of
the equation $Lu(x)=f(x)$) is the kernel of the integral operator inverse to
$L$. As the kernel of the unit operator is the Dirac delta\ndash function
$\delta^4(x'-x)$, the Green function is a fundamental solution of the
non\ndash homogeneous equation
	\begin{equation}	\label{9.a}
Lu(x)=f(x) ,
	\end{equation}
\ie treated as a generalized function $g(x',x)$ is a solution of
	\begin{equation}	\label{9.b}
L_{x'}g(x',x) = \delta^4(x'-x) .
	\end{equation}
Given a Geen function $g(x',x)$, the solution of~\eref{9.a} is
	\begin{equation}	\label{9.c}
u(x') = \int g(x',x)f(x) \,\od^4 x.
	\end{equation}

	A concrete Green function $g(x',x)$ for $L$ (or~\eref{9.a})
satisfies, besides~\eref{9.b}, certain (homogeneous) boundary conditions on
$x'$ with fixed $x$, \ie it is the solution of a fixed boundary\ndash value
problem for equation~\eref{9.b}. Hence, if $g_f(x',x)$ is some fundamental
solution, then
	\begin{equation}	\label{9.d}
g(x',x ) = g_f(x',x) + g_0(x',x)
	\end{equation}
where $g_0(x',x )$ is a solution of the homogeneous equation
$L_{x'}g_0(x',x ) = 0 $ chosen such that  $g(x',x )$ satisfies the required
boundary conditions.

	Suppose $g(x',x )$ is a Green function of $L$ for some boundary\ndash
value (or initial\ndash value) problem. Then, using~\eref{9.b}, we can verify
that
	\begin{equation}	\label{9.e}
L_{x'}\Bigl(\int g(x',x )u(x) \,\od^3\Vect{x}\Bigr)
= \delta(ct'-ct)u(x')
	\end{equation}
where $x=(ct,\Vect{x})$ and $x'=(ct',\Vect{x}')$. Therefore the
solution of the problem
	\begin{equation}	\label{9.f}
L_xu(x) = 0 , \quad u(ct_0,\Vect{x}) = u_0(\Vect{x})
	\end{equation}
is
	\begin{equation}	\label{9.g}
u(x) = \int g\bigl(x,(ct_0,\Vect{x}_0)\bigr)  u_0(\Vect{x}_0)
	\,\od^3 \Vect{x}_0,
\quad
x_0 = (ct_0,\Vect{x}_0)
\qquad\text{ for $t\not=t_0$} .
	\end{equation}
	From here we can make the conclusion that, if~\eref{9.f} admits an
evolution operator $\ope{U}$ such that (cf.~\eref{2.2})
	\begin{equation}	\label{9.h}
u(x) \equiv u(ct,\Vect{x})
=\ope{U}(t,t_0)u(ct_0,\Vect{x}) ,
	\end{equation}
then the r.h.s\ of~\eref{9.g} realizes $\ope{U}$ as an integral operator with
a kernel equal to the Green function $g$.

	Since all (relativistic or not) wave equations are versions
of~\eref{9.f}, the corresponding evolution operators, if any, and Green
functions (propagators) are connected as just described. Moreover, if some
wave equation does not admit (directly) evolution operator, e.g.\ if it is
of order greater than one, then we can \emph{define} it as the corresponding
version of the integral operator in the r.h.s.\ of~\eref{9.g}. In this way
is established a one\nobreakdash-to\nobreakdash-one onto correspondence
between the evolution operators and Green functions for any particular
problem like~\eref{9.f}.

	And a last general remark. The so-called S-matrix finds a lot of
applications in quantum theory
~\cite{Bjorken&Drell-1,Itzykson&Zuber,Bogolyubov&Shirkov}. By definition $S$
is an operator transforming the system's state vector
$\psi(-\infty,\Vect{x})$ before scattering (reaction) into the one
$\psi(+\infty,\Vect{x})$ after it:
\[
\lim_{t\to+\infty}\psi(ct,\Vect{x})
=: S \lim_{t\to-\infty}\psi(ct,\Vect{x}) .
\]
So, e.g.\ when~\eref{9.h} takes place, we have
	\begin{equation}	\label{9.j}
S
=  \lim_{t_\pm \to \pm\infty} \ope{U}(t_+,t_-)
=: \ope{U}(+\infty,-\infty) .
	\end{equation}

	Thus the above-mentioned connection between evolution operators and
Green functions can be used for expressing the S\ndash matrix in terms of
propagators. Such kind of formulae are often used in relativistic quantum
mechanics~\cite{Bjorken&Drell-1}.

\subsection {Nonrelativistic case (Schr\"odinger equation)}
\label{Subsect9.1}

	The (retarded) Green function $g(x',x)$, $x',x\in M$, for the
Schr\"odinger equation~\eref{2.1} is defined as the solution of the
boundary\ndash value problem~\cite[\S~22]{Bjorken&Drell-1}
	\begin{align}	\label{9.1}
\bigl[ \ih\frac{\pd}{\pd t'} - \Ham(x') \bigr] g(x',x) &= \delta^4(x'-x),
\\
\label{9.2}
g(x',x) &= 0 \qquad \text{for $t'<t$}
	\end{align}
where $x'=(ct',\Vect{x}')$ $x=(ct,\Vect{x})$,
$\Ham(x)$ is system's Hamiltonian, and $\delta^4(x'-x)$ is the
4-dimensional (Dirac) $\delta$-function.

	Given $g$, the solution  $\psi(x')$ (for $t'>t$) of~\eref{2.1} is%
\footnote{%
One should not confuse the notation $\psi(x)=\psi(ct,\Vect{x})$,
$x=(ct,\Vect{x})$ of this section and $\psi(t)$ from Sect.~\ref{Sect2}.
The latter is the wavefunction at a moment $t$ and the former is its value at
the spacetime point $x=(ct,\Vect{x})$. Analogously,
\(
\Psi_\gamma(x)\equiv\Psi_\gamma(ct,\Vect{x})
:=l_{\gamma(t)}^{-1}\bigl(\psi(ct,\Vect{x})\bigr)
\)
 should not be confused with $\Psi_\gamma(t)$ from Sect.~\ref{Sect2}. A
notation like $\psi(t)$ and $\Psi_\gamma(t)$ will be used if the spatial
parts of the arguments are inessential, as in Sect.~\ref{Sect2}, and there is
no risk of ambiguities.%
}
	\begin{equation}	\label{9.3}
\theta(t'-t)\psi(x')
= \ih\int\ordinary^3\Vect{x} g(x',x)\psi(x)
	\end{equation}
where the $\theta$-function $\theta(s),\ s\in\mathbb{R}$, is defined by
$\theta(s)=1$ for $s>0$ and $\theta(s)=0$ for $s<0$.

	Combining~\eref{2.2} and~\eref{9.3}, we find the basic connection
between the evolution operator $\ope{U}$ and Green function of Schr\"odinger
equation:
	\begin{equation}	\label{9.4}
\theta(t'-t)\bigl[\ope{U}(t',t)\bigl(\psi(ct,\Vect{x}')\bigr)\bigr]
= \ih\int\ordinary^3\Vect{x}
g\bigl( (ct',\Vect{x}'),(ct,\Vect{x}) \bigr)
\psi(ct,\Vect{x}) .
	\end{equation}
Actually this formula, if $g$ is known, determines $\ope{U}(t',t)$ for all
$t'$ and $t$, not only for $t'>t$, as $\ope{U}(t',t)=\ope{U}^{-1}(t,t')$ and
$\ope{U}(t,t)=\id_\fibre$ with $\fibre$ being the system's Hilbert space
(see~\eref{2.2} or~\cite[sect.~2]{bp-BQM-introduction+transport}).
Consequently \emph{the evolution operator for the Schr\"odinger equation
can be represented as an integral operator whose kernel, up to the
constant $\ih$, is exactly the (retarded) Green function for it}.

	To write the bundle version of~\eref{9.4}, we introduce the
\emph{Green operator} which is simply a multiplication with the Green
function:
	\begin{equation}	\label{9.5}
\ope{G}(x',x) := g(x',x)\id_\fibre\colon\fibre\to\fibre.
	\end{equation}
The corresponding to it \emph{Green morphism} $G$ is given via
(see~\eref{2.11} and cf.~\eref{2.4})
	\begin{equation}	\label{9.6}
\mor{G}_\gamma(x',x)
:= l_{\gamma(t')}^{-1} \circ\ope{G}(x',x)\circ l_{\gamma(t)}
=  g(x',x)l_{\gamma(t')}^{-1}\circ l_{\gamma(t)}
\colon\fibreover{\gamma(t)}\to\fibreover{\gamma(t')} .
	\end{equation}

	Now, acting on~\eref{9.4} from the left by $l_{\gamma(t')}^{-1}$ and
using~\eref{2.3} and~\eref{2.4}, we obtain
	\begin{equation}	\label{9.7}
\theta(t'-t)
\bigl[\mor{U}_\gamma(t',t)\bigl(\Psi_\gamma(ct,\Vect{x}')\bigr)\bigr]
= \ih\int\ordinary^3\Vect{x}
\mor{G}_\gamma\bigl( (ct',\Vect{x}'),(ct,\Vect{x}) \bigr)
\Psi_\gamma(ct,\Vect{x}) .
	\end{equation}
Therefore for the Schr\"odinger equation the
\emph{%
	evolution transport $\mor{U}$ can be represented as an integral
	operator with kernel equal to $\ih$ times the Green morphism $G$%
}.

	Taking as a starting point~\eref{9.4} and~\eref{9.7}, we can obtain
different representations for the evolution operator and transport by applying
concrete formulae for the Green function. For example, if a complete set
$\{\psi_a(x)\}$ of orthonormal solutions of Schr\"odinger equation satisfying
the completeness condition%
\footnote{%
Here and below the symbol $\sum_{a}$ denotes a sum and/or integral over the
discrete and/or continuous spectrum. The asterisk ($*$) means complex
conjugation.%
}
\[
  \sum_a \psi_a(ct,\Vect{x}')\psi_a^*(ct,\Vect{x})
=\delta^3(\Vect{x}'- \Vect{x})
\]
is know, then~\cite[\S~22]{Bjorken&Drell-1}
\[
g(x',x)
= \iih\theta(t'-t) \sum_a \psi_a(x')\psi_a^*(x)
\]
which, when substituted into~\eref{9.4}, implies
	\begin{equation}	\label{9.8}
\ope{U}(t',t)\psi(ct,\Vect{x}')
=\sum_a \psi_a(x')
\int\od^3\Vect{x}\,\psi_a^*(ct,\Vect{x}) \psi(ct,\Vect{x}).
	\end{equation}
Note, the integral in this equation is equal to the  $a$-th coefficient
of the expansion of $\psi$ over $\{\psi_a\}$.

\subsection {Dirac equation}
\label{Subsect9.2}

	Since the Dirac equation~\eref{7.1} is a first-order linear partial
differential equation, it admits both evolution operator and Green
function(s) (propagator(s)). From a generic view\nobreakdash-point, the
only difference from the Schr\"odinger equation is that~\eref{7.1} is a
\emph{matrix} equation; so the corresponding Green functions are actually
Green matrices, \ie  Green matrix\ndash valued functions. Otherwise the
results of Subsect.~\ref{Subsect9.1} are \emph{mutatis mutandis} applicable
to the theory of Dirac equation.

	The (retarded) Green matrix (function) or propagator for Dirac
equation~\eref{7.1} is a $4\times4$ matrix\ndash valued function $g(x',x)$
depending on two arguments $x',x\in M$ and such
that~\cite[sect.~2.5.1 and~2.5.2]{Itzykson&Zuber}
	\begin{align}
\label{9.20}
(\ih \Slashed{D}_{x'} - mc\openone_4) g(x',x) &= \delta(x'-x)
\\
\label{9.21}
g(x',x) &= 0 \qquad \text{for $t<t'$}.
        \end{align}

	For a free Dirac particle, \ie for $\Slashed{D}=\slashed{\pd}$ or
$\ope{A}_\mu=0$, the explicit expression $g_0(x',x)$ for $g(x',x)$ is
derived in~\cite[sect.~2.5.1]{Itzykson&Zuber}, where the notation
$\mathcal{K}$ instead of $g_0$ is used. In an external electromagnetic field
$\ope{A}_\mu$ the Green matrix $g$ is a solution of the integral equation%
\footnote{%
The derivation of~\eref{9.22} is the same as for the Feynman propagators
$S_F$ and $S_A$ given in~\cite[sect.~2.5.2]{Itzykson&Zuber}. The propagators
$S_F$ and $S_A$ correspond to $g_0$ and $g$ respectively, but satisfy other
boundary conditions~\cite{Itzykson&Zuber,Bjorken&Drell-1}.%
}
	\begin{equation}	\label{9.22}
g(x',x)
= g_0(x',x) + \int\od^4y\, g_0(x',y) \frac{e}{c}\Slashed{\ope{A}}(y) g(y,x)
	\end{equation}
which includes the corresponding boundary condition.%
\footnote{%
Due to~\eref{9.d}, the integral equation~\eref{9.22} is valid for any Green
function (matrix) of the Dirac equation~\eref{7.1}.%
}
The iteration of this equation results in the perturbation series for $g$
(cf.~\cite[sect.~2.5.2]{Itzykson&Zuber}).

	If the (retarded) Green matrix $g(x',x)$ is known, the solution
$\psi(x')$ of Dirac equation (for $t'>t$) is
	\begin{equation}	\label{9.23}
\theta(t'-t)\psi(x')
= \ih\int\od^3\Vect{x}\, g(x',x)\gamma^0\psi(x) .
	\end{equation}

	Hence, denoting by $\ope{U}$ the non\ndash relativistic (see
Subsect.~\ref{Subsect3.4}) Dirac evolution operator, from the
equations~\eref{2.2}, and~\eref{9.23}, we find:
	\begin{equation}	\label{9.24}
\theta(t'-t)
\bigl[ \ope{U}(t',t)\bigl(\psi(ct,\Vect{x}') \bigr)\bigr]
= \ih\int\od^3\Vect{x}\,
	g\bigl((ct',\Vect{x}'),(ct,\Vect{x})\bigr)
	\gamma^0\psi(ct,\Vect{x}) .
	\end{equation}
So, the evolution operator admits an integral representation
whose kernel, up to the right multiplication with  $\ih\gamma^0$, is equal to
the (retarded) Green function for Dirac equation.

	Similarly to~\eref{9.7}, now the bundle version of~\eref{9.24} is
	\begin{align}
		\label{9.25a}
\theta(t'-t)
\bigl[ \mor{U}(t',t)
	\bigl(\Psi_\gamma(ct,\Vect{x}')\bigr) \bigr]
& = \ih\int\od^3\Vect{x}\,
	\mor{G}_\gamma(x',x) G^0(\gamma(t))\Psi_\gamma(ct,\Vect{x}),
	\end{align}
where $G^0(x)$ is defined by~\eref{7.17} with $\mu=0$ and (cf.~\eref{9.6})
	\begin{equation}	\label{9.26}
\mor{G}_\gamma(x',x)
:= l_{\gamma(t')}^{-1}\circ \ope{G}(x',x) \circ l_{\gamma(t)},
\quad
\mor{G}(x',x)
:= l_{x'}^{-1}\circ \ope{G}(x',x) \circ l_{x}
	\end{equation}
with (\cf.~\eref{9.5})
	\begin{equation}	\label{9.27}
\ope{G}(x',x) = g(x',x) \id_\fibre .
	\end{equation}
(Here $\fibre$ is the space of 4-spinors.)

	Analogously to the above results, one can obtained such for other
propagators, \eg for the Feynman one~\cite{Bjorken&Drell-1,Itzykson&Zuber},
but we are not going to do this here as it is a trivial variant of the
procedure described.

\subsection {Klein-Gordon equation}
\label{Subsect9.3}

	The (retarded) Green function $g(x',x)$ for the Klein-Gordon
equation~\eref{8.1} is a solution to the boundary\ndash value
problem~\cite[sect.~1.3.1]{Itzykson&Zuber}
	\begin{align}
\label{9.31}
\Bigl.\Bigl(
\mathcal{D}^\mu\mathcal{D}_\mu + \frac{m^2c^2}{\hbar^2}
	\Bigr)\Bigr|_{x'} g(x',x)
&= \delta^4(x'-x),
\\
\label{9.32}
g(x',x) &= 0 \qquad \text{for $t'<t$}.
	\end{align}
For a free particle its explicit form can be found
in~\cite[sect.~1.3.1]{Itzykson&Zuber}.

	A simple verification proves that, if $g(x',x)$ is known, the
solution $\phi$ of~\eref{8.1} (for $t'>t$) is given by ($x^0=ct$)
	\begin{equation}	\label{9.33}
\theta(t'-t)\phi(x')
= \mspace{-1.23mu}
  \int\od^3\Vect{x}\,
	\Bigl[ \frac{\pd g(x',x)}{\pd x^0} \phi(x)
	+ g(x',x) \Bigl( 2\frac{\pd\phi(x)}{\pd x^0}
		-\frac{e}{\ih c}\ope{A}^0(x)\phi(x) \Bigr)
	\Bigr] .
	\end{equation}

	Introducing the matrices
	\begin{equation}	\label{9.34}
\psi(x)
:= 	\begin{pmatrix} \phi(x) \\ \pd_0|_x\phi(x) \end{pmatrix},
\qquad
\mathsf{g}(x',x)
:=	\begin{pmatrix} \mathcal{D}_0|_x g(x',x) \\ 2g(x',x) \end{pmatrix},
	\end{equation}
we can rewrite~\eref{9.33} as
	\begin{equation}	\label{9.35}
\theta(t'-t)\phi(x')
= \int\od^3\Vect{x}\,
	\mathsf{g}^\top(x',x)\cdot\psi(x)
	\end{equation}
where the dot ($\cdot$) denotes matrix multiplication.

	An important observation is that for $\psi$  the Klein-Gordon equation
transforms into first\ndash order Schr\"odinger\ndash type equation (see
Subsect.~\ref{Subsect3.5}) with Ha\-miltonian
$\lindexrm[\Ham]{\mspace{32mu}c}{K-G}$ given by~\eref{5.3} in which
$\id_{ldots}$ is replaced by $c\id_{ldots}$.

	Denoting the (retarded) Green function, which is in fact $2\times2$
matrix, and the evolution operator for this equation by
\(
\widetilde{\ope{G}}(x',x) =
\bigl[ \widetilde{\ope{G}}_{~b}^{a}(x',x)\bigr]_{a,b=1}^{2}
\)
and
\(
\widetilde{\ope{U}}(t',t) =
\bigl[ \widetilde{\ope{U}}_{~b}^{a}(t',t)\bigr]_{a,b=1}^{2},
\)
respectively, we see that (\cf Subsect.~\ref{Subsect9.1},
equation~\eref{9.3})
	\begin{equation}	\label{9.36}
\theta(t'-t) \bigl[\widetilde{\ope{U}}(t',t) \psi(ct,\Vect{x})\bigr]
= \theta(t'-t)\psi(x')
= \int\od^3\Vect{x}\,
	\widetilde{\ope{G}}(x',x) \psi(x) .
	\end{equation}
Comparing this equations with~\eref{9.34} and~\eref{9.35}, we find
\[
\bigl(
\widetilde{\ope{G}}_{~1}^{1}(x',x),\widetilde{\ope{G}}_{~2}^{1}(x',x) \bigr)
= \mathsf{g}^\top(x',x) .
\]
The other matrix elements of $\widetilde{\ope{G}}$ can also be connected with
$\mathsf{g}(x',x)$ and its derivatives, but this is inessential for the
following.

	In this way we have connected, via~\eref{9.36}, the evolution
operator and the (retarded) Green function for a concrete first\ndash order
realization of Klein\ndash Gordon equation. It is almost evident that this
procedure \emph{mutatis mutandis} works for any such realization; in every
case the corresponding Green function (resp.\ matrix) being a (resp.\
matrix\ndash valued) function of the Green function $g(x',x)$ introduce
via~\eref{9.31} and~\eref{9.32}. For instance, the treatment of the 5\ndash
dimensional realization given by~\eref{8.2} and~\eref{8.3} is practically
identical to the one of Dirac equation in Subsect.~\ref{Subsect9.2}, only the
$\gamma$\ndash matrices $\gamma^\mu$ have to be replace with the $5\times5$
matrices $\Gamma^\mu$ (defined by~\eref{8.2}). This results in a $5\times5$
matrix evolution operator $\ope{U}(x',x)$, etc.

	Since the bundle version of~\eref{9.36} or an analogous result for
$\ope{U}(x',x)$ is absolutely trivial (\cf Subsect.~\ref{Subsect9.1}
and~\ref{Subsect9.2} resp.), we are not going to write it here; up to the
meaning of notation it coincides with~\eref{9.7} or~\eref{9.25a}
respectively.

\section {Conclusion}
\label{Conclusion}

	In this investigation we have reformulated the relativistic wave
equations in terms of fibre bundles. In the bundle formulation the
wavefunctions are represented as (state) liftings of paths
(equivalently: sections along paths) (time\ndash dependent approach) or
simply sections (covariant approach) of a suitable vector bundle over the
spacetime.  The covariant approach has an advantage of being explicitly
covariant while in the time\ndash dependent one the time plays a privilege
r\^ole. In both cases the evolution (in time or in spacetime resp.) is
described via a linear transport in the bundle mentioned.
The state liftings/sections are linearly transported by means of the
corresponding (evolution) transports. We also have explored some links between
evolution operators or transports and the retarded Green functions (or
matrices) for the corresponding wave equations: the former turn to have
realization as integral operators whose kernel is equal to the latter ones up
to a multiplication with a constant complex number or matrix.

	These connections suggest the idea for introducing `retarded', or, in
a sense, `causal' evolution operators or transports as a product of the
evolution operators or transports with $\theta$\ndash function of the
difference of the times corresponding to the first and second arguments of the
transport or operator.

	Most of the possible generalizations of the bundle non\ndash
relativistic quantum mechanics, pointed
in~\cite{bp-BQM-interpretation+discussion}, are \emph{mutatis mutandis} valid
with respect to the bundle version of relativistic quantum mechanics,
developed in the present work. The only essential change is that, in the
relativistic region, the spacetime model is fixed as the Minkowski spacetime.

	A further development of the ideas presented in this paper leads to
their application to (quantum) field theory which will be done elsewhere.


	We shall end with some comments on the material of
 Subsect.~\ref{Subsect3.3}.
As we saw there, the fibre bundle formalism developed for the solutions
of Schr\"odinger equation can successfully be applied for the solutions of
(systems of) linear ordinary differential equations. For this purpose the
system of equations, if they are of order greater than one, has to be
transformed into a system of first-order equations. It can always be written
in a Schr\"odinger\ndash{like} form~\eref{3.1} to which the developed
in~\cite{bp-BQM-introduction+transport,bp-BQM-equations+observables,
	bp-BQM-pictures+integrals,bp-BQM-mixed_states+curvature,
	bp-BQM-interpretation+discussion}
bundle approach can be applied \textit{mutatis mutandis}.

	Therefore, in particular, to a system of linear ordinary (with respect
to `time') differential equations corresponds a suitable linear transport
along paths in an appropriately chosen fibre bundle. As most of the
fundamental equations of physics are expressed by such systems of equations,
they admit fibre bundle (re)formulation analogous to the one of
Schr\"odinger equation.

	An interesting consequence of this discussion is worth mentioning.
Suppose a system of the above-described type is an Euler-Lagrange (system of)
equation(s) for some Lagrangian. Applying the outlined `bundle' procedure,
we see that to this Lagrangian corresponds some (evolution) linear transport
along paths in a suitable fibre bundle. The bundle and the transport are
practically (up to isomorphisms) unique if the Euler-Lagrange equations are of
first order with respect to time. Otherwise there are different (but
equivalent) such objects corresponding to the given Lagrangian. Hence, some
Lagrangians admit description in terms of linear transports along paths. In
more details the correspondence between Lagrangians (or Hamiltonians) and
linear transports along paths will be explored elsewhere.


\appendix
\renewcommand{\thesection}{Appendix~\Alph{section}}

\section
[\hspace*{4.75em}Linear transports along maps in fibre bundles]
{Linear transports along maps in\\ fibre bundles}

	\renewcommand{\thesection}{\Alph{section}}
	\label{AppendixA}

	In this appendix we recall a few simple facts concerning (linear)
transports along maps, in particular along paths, required for the present
investigation. The below\ndash presented material is abstracted
from~\cite{bp-TM-general,bp-LTP-general,bp-normalF-LTP} where further details can be found
(see also~\cite[sect.~3]{bp-BQM-introduction+transport}).

	Let $(E,\pi,B)$ be a topological bundle with base $B$,
bundle (total, fibre) space $E$, projection $\pi:E\to B$, and homeomorphic
fibres $\pi^{-1}(x),\ x\in B$.  Let the set $N$ be not empty,
$N\neq\varnothing$, and there  be given a map $\varkappa:N\to B$. By $\id_X$
is denoted the identity map of a set $X$.

	\begin{Defn}	\label{DefnLT.1}
	A transport along maps in the bundle $(E,\pi,B)$ is a map
$K$ assigning to any map $\varkappa:N\to B$ a map  $K^\varkappa$,
transport along $\varkappa$, such that
$K^\varkappa:(l,m)\mapsto K_{l\to m}^{\varkappa}$, where for
every $l,m\in N$ the map
\begin{equation}	\label{LT.1}
K_{l\to m}^{\varkappa}:\pi^{-1}(\varkappa(l))\to \pi^{-1}(\varkappa(m)),
\end{equation}
called transport along  $\varkappa$ from  $l$ to $m$, satisfies the
equalities:
\begin{align}
K_{m\to n}^{\varkappa} \circ  K_{l\to m}^{\varkappa} &=
	K_{l\to n}^{\varkappa} , & l,m,n\in N, 		\label{LT.2}\\
K_{l\to l}^{\varkappa} &= \id_{\pi^{-1}(\varkappa(l))},
				 & l\in N.		\label{LT.3}
\end{align}
\end{Defn}

	If $(E,\pi,B)$ is a complex (or real) vector bundle and the
maps~\eref{LT.1} are linear, \ie
	\begin{equation}      \label{LT.4}  \!
K_{l\to m}^{\varkappa}(\lambda u + \mu v) =
\lambda K_{l\to m}^{\varkappa}u + \mu K_{l\to m}^{\varkappa}v,\ \>
\lambda,\mu\in\mathbb{C\mathrm{\ (or\ }R)},\ \>
u,v\in\pi^{-1}(\varkappa(l)),
	\end{equation}
the transport $K$ is called \emph{linear}. If $\varkappa$ belongs to the set
of paths in $B$,
\(
\varkappa\in\{\gamma\colon J\to B,\
	J\text{ being }\mathbb{R}\text{-interval}\},
\)
we said that the transport $K$ is \emph{along paths}.

	For the present work is important that the class of linear transports
along the identity map $\id_B$ of $B$ coincides with the class of \emph{flat}
linear transports along paths%
\footnote{%
The \emph{flat} linear transports along paths are defined as ones with
vanishing curvature operator~\cite[sect.~2]{bp-LTP-Cur+Tor-prop}.
By~\cite[theorem.~6.1]{bp-LTP-Cur+Tor-prop} we can equivalently define them by
the property that they depend only on their initial and final points, \ie if
$K_{s\to t}^{\gamma}$, $\gamma\colon J\to B$, $s,t\in J\subset\mathbb{R}$,
depends only on $\gamma(s)$ and $\gamma(t)$ but not on the path $\gamma$
itself.%
}
(see the comments after equation~(2.3) of~\cite{bp-TM-general}).

	The general form of a transport along maps is described by the
following result.
	\begin{Thm}	\label{ThmLT.1}
Let $\varkappa:N\to B$. The map
$K:\varkappa\mapsto K^\varkappa:(l,m)\mapsto K_{l\to m}^{\varkappa}$,
$l,m\in N$
is a transport along $\varkappa$ if and only if there exist a set
$Q$ and a family of bijective maps
 $\{ F_{n}^{\varkappa}:\pi^{-1}(\varkappa(n))\to Q,\ n\in N \}$
such that
	\begin{equation}	\label{LT.5}
K_{l\to m}^{\varkappa} = \left(F_{m}^{\varkappa}\right)^{-1} \circ
\left(F_{l}^{\varkappa}\right),	\quad l,m\in N.
	\end{equation}
The maps $F_{n}^{\varkappa}$ are defined up to a left composition with
bijective map depending only on $\varkappa$, i.e.~\eref{LT.5} holds for given
families of maps
$\{ F_{n}^{\varkappa}:\pi^{-1}(\varkappa(n))\to Q,\ n\in N \}$ and
\(
\{ ^\prime\! F_{n}^{\varkappa}:\pi^{-1}(\varkappa(n))\to
{}^\prime\! Q,\ n\in N \}
\)
for some sets $Q$ and  $^\prime\! Q$ iff there is a bijective map
$D^\varkappa:Q\to{}^\prime\! Q$ such that
	\begin{equation}	\label{LT.6}
^\prime\! F_{n}^{\varkappa} = D^\varkappa \circ F_n^\varkappa,\quad n\in N.
	\end{equation}
	\end{Thm}

	For the purposes of this investigation we need a slight
generalization of~\cite[definition~4.1]{bp-TM-general}, viz. we want to
replace in it $\mathbf{N}\subset\mathbb{R}^k$ with an arbitrary differentiable
manifold. Let $N$ be a differentiable manifold and
$\{x^a\ :\ a=1,\dots,\dim{N}\}$ be coordinate system in a neighborhood of
$l\in N$. For
$\varepsilon\in(-\delta,\delta)\subset\mathbb{R}$, $\delta\in\mathbb{R}_+$
and  $l\in N$ with coordinates $l^a=x^a(l)$, we define
$l_b(\varepsilon)\in N$, $b=1,\dots,\dim{N}$ by
\(
l_{b}^{a}(\varepsilon):=x^a(l_b(\varepsilon)):=l^a +
	\varepsilon \delta_{b}^{a}
\)
where the Kroneker $\delta$\ndash symbol is given by $\delta_{b}^{a}=0$ for
$a\not=b$ and $\delta_{b}^{a}=1$ for $a=b$.
	Let $\xi=(E,\pi,B)$ be a vector bundle,
$\varkappa\colon N\to B$ be injective (\ie 1:1 mapping),
and $\mathrm{Sec}^p(\xi)$ (resp.\ $\mathrm{Sec}(\xi)$) be the
set of $C^p$ (resp.\ all) sections over $\xi$.
Let $L_{l\to m}^{\varkappa}$ be a $C^1$ (on $l$) \emph{linear} transport along
$\varkappa$.
Now the modified definition reads:%
\footnote{%
We present below, in definition~\ref{DefnLT.2}, directly the definition of a
\emph{section}\ndash derivation along injective mapping $\varkappa$ as only
it will be employed in the present paper (for $\varkappa=\id_N$). If
$\varkappa$ is not injective, the mapping~\eref{LT.7} could be
multiple\ndash valued at the points of self\ndash intersection of
$\varkappa$, if any. For some details when $\varkappa$ is an arbitrary path
in $N$, see~\cite[subsect.~3.3]{bp-BQM-introduction+transport}.%
}
	\begin{Defn}	\label{DefnLT.2}
	The $a$-th, $1\le a\le\dim N$, partial
(section\ndash)derivation along maps generated by  $L$ is a  map
${_a  }\mathcal{D}:\varkappa\mapsto{_a  }\mathcal{D}^\varkappa$
where the $a$-th (partial) derivation ${_a  }\mathcal{D}^\varkappa$
along $\varkappa$ (generated by $L$)  is  a map
	\begin{equation}	\label{LT.7}
{_a  }\mathcal{D}^\varkappa : l \mapsto \mathcal{D}^\varkappa_{l^a} ,
	\end{equation}
where the $a$-th (partial) derivative $\mathcal{D}^\varkappa_{l^a}$ along
$\varkappa$ at $l$ is a map
	\begin{equation}	\label{LT.9}
	\mathcal{D}_{l^a}^{\varkappa} :
\mathrm{Sec}^1\left(\left.\xi\right|_{\varkappa(N)}\right) \to
\pi^{-1}(\varkappa(l))
	\end{equation}
defined for
$\sigma\in\mathrm{Sec}^1\left(\xi|_{\varkappa(N)}\right)$ by
	\begin{equation} 	\label{LT.8}
\left({_a  }\mathcal{D}^\varkappa\sigma\right) (\varkappa(l)) :=
\lim_{\varepsilon\to 0}  	\left[ \frac{1}{\varepsilon} \bigl(
L_{l_a(\varepsilon) \to l}^{\varkappa}\sigma(\varkappa(l_a(\varepsilon)))
				- \sigma(\varkappa(l)) \bigr) \right].
	\end{equation}
	\end{Defn}

	Accordingly can be modified the other definitions
of~\cite[sect.~4]{bp-TM-general}, all the results of it being
\emph{mutatis mutandis} valid. In particular, we have:
	\begin{Prop}	\label{PropLT.1}
	The operators $\mathcal{D}^\varkappa_{l^a}$ are
($\mathbb{C}$-)linear
and
	\begin{equation}	\label{LT.10}
\mathcal{D}_{m^a}^{\varkappa} \circ L_{l\to m}^{\varkappa} \equiv 0 .
	\end{equation}
	\end{Prop}

	\begin{Prop}	\label{PropLT.2}
	If
$\sigma\in\mathrm{Sec}^1\left(\xi|_{\varkappa(N)}\right)$,
then
	\begin{equation}	\label{LT.11}
\mathcal{D}_{l^a}^{\varkappa} \sigma
= \sum_i
\left[
	\frac{\partial\sigma^i(\varkappa(l))}{\partial l^a}
	+ \sum_{j} {_a  }\Gamma_{\;j}^{i}(l;\varkappa)\sigma^j(\varkappa(l))
\right] e_i(l) ,
	\end{equation}
where $\{e_i(l)\}$ is a basis in $\pi^{-1}(\varkappa(l))$,
 $\sigma(\varkappa(l))=:\sum_{i}\sigma^i(\varkappa(l)) e_i(l)$,
and the \textbf{coefficients} of  $L$ are defined by
	\begin{equation}	\label{LT.12}
{_a  }\Gamma_{\;j}^{i}(l;\varkappa) :=
\left.
\frac{\partial L_{\;j}^{i}(l,m;\varkappa)}{\partial m^a}
\right|_{m=l} =
- \left.
\frac{\partial L_{\;j}^{i}(m,l;\varkappa)}{\partial m^a}
\right|_{m=l}.
	\end{equation}
Here $L_{\;i}^{j}(\cdots)$ are the components of $L$,
$L_{l\to m}^{\varkappa}e_i(l)=:\sum_j L_{\;i}^{j}(m,l;\varkappa)e_j(m)$.
	\end{Prop}

	The above general definitions and results will be used in this work
in the special cases of linear transports along paths and linear transports
along the identity map of the bundle's base.

\renewcommand{\thesection}{Appendix~\Alph{section}}

\section[\hspace*{4.75em}Matrix operators]{Matrix operators}

	\renewcommand{\thesection}{\Alph{section}}
	\label{AppendixB}

	In this appendix we point to some peculiarities of linear
(matrix) operators acting on $n\times1$, $n\in\mathbb{N}$, matrix fields
over the space\nobreakdash-time $\bbase$. Such operators appear naturally in
the theory of Dirac equation where one often meets $4\times4$ matrices whose
elements are operators; e.g.\ an operator of this kind is
\(
\slashed{\partial} := \gamma^\mu\partial_\mu =
\bigl[ (\gamma^\mu)_{~\beta}^{\alpha}\partial_\mu \bigr]_{\alpha,\beta=0}^{3}
\)
where $\gamma^\mu$ are the well known Dirac $\gamma$\ndash
matrices~\cite{Itzykson&Zuber,Bjorken&Drell-1}.

	We call an $n\times n$, $n\in\mathbb{N}$ matrix
$B=[b_{~\beta}^{\alpha}]_{\alpha,\beta=1}^{n}$
a \emph{(linear) matrix operator}%
\footnote{%
Also a good term for such an object is (linear) \emph{matrixor}.%
}
if $b_{~\beta}^{\alpha}$ are (linear) operators acting on the space
$\mathit{K}^1$ of $C^1$ functions $f\colon M\to\mathbb{C}$.
If $\{ f_{\nu}^{0} \}$ is a basis in the set $M(n,1)$ of $n\times1$ matrices
with the $\mu$\nobreakdash-th element of $f_{\nu}^{0}$ being
$(f_{\nu}^{0})^\mu:=\delta_{\nu}^{\mu}$,%
\footnote{%
$\delta_{\nu}^{\mu}=0$ for $\mu\not=\nu$ and
$\delta_{\nu}^{\mu}=1$ for $\mu=\nu$. From here to equation~\eref{mo.9} in
this appendix the Greek indices run from 1 to $n\ge1$.%
}
then by definition
	\begin{equation}	\label{mo.1}
B\psi :=B(\psi) := B\cdot(\psi) :=
\sum_{\alpha,\beta=1}^{n}
\bigl( b_{~\beta}^{\alpha} (\psi_{0}^{\beta} \bigr)  f_{\alpha}^{0},
\qquad
\psi = \psi_{0}^{\beta} f_{\beta}^{0} \in M(n,1).
	\end{equation}
For instance, we have
\(
\slashed{\pd}\psi
= \sum_{\alpha,\beta,\mu=0}^{3}
  (\gamma^\mu)_{~\beta}^{\alpha}(\pd_\mu\psi_{0}^{\beta}) f_{\alpha}^{0}.
\)

	To any constant matrix $C=[c_{~\beta}^{\alpha}]$,
$c_{~\beta}^{\alpha}\in\mathbb{C}$, corresponds a matrix operator
 $\overline{C}:=[c_{~\beta}^{\alpha}\id_{\mathit{K}^1}]$. Since
$C\psi\equiv\overline{C}\psi$ for any $\psi$, we identify $C$ and
$\overline{C}$ and will make no difference between them.

	The multiplication of matrix operators, denoted by $\odot$, is a
combination of matrix multiplication (denoted by $\cdot$) and maps
(operators) composition (denoted by $\circ$). If $A=[a_{~\beta}^{\alpha}]$
and $B=[b_{~\beta}^{\alpha}]$ are matrix operators, the product of $A$
and $B$ is also a matrix operator such that

	\begin{equation}	\label{mo.2}
AB := A\odot B :=
\biggl[ \sum_{\mu}^{} a_{~\mu}^{\alpha} \circ b_{~\beta}^{\mu} \biggr] .
	\end{equation}
	One can easily show that this is an associative operation. It is
linear in the first argument and, if its first argument is linear matrix
operator, then it is linear in its second argument too. For constant matrices
(see above), the multiplication~\eref{mo.2} coincides with the usual matrix
one.

	Let $\{f_{\alpha}(x)\}$ be a basis in $M(n,1)$ depending on $x\in M$
and $f(x):=[f_{\alpha}^{\beta}(x)]$ be defined by the expansion
$f_{\alpha}(x)=f_{\alpha}^{\beta}(x)f_{\beta}^{0}$.
The \emph{matrix of a matrix operator} $B=[b_{~\beta}^{\alpha}]$ with respect
to $\{f_\alpha\}$, \ie the matrix of the matrix elements of $B$ considered as
an operator, is also a matrix operator
$\Mat{B}:=[B_{~\beta}^{\alpha}]$  such that
	\begin{equation}	\label{mo.3}
B\psi|_x
=: \sum_{\alpha,\beta}^{}
    \bigl( B_{~\beta}^{\alpha}(\psi^\beta) \bigr) \bigl.\bigr|_x f_\alpha(x),
\qquad
\psi(x) = \psi^\beta(x)f_\beta(x).
        \end{equation}
Therefore
	\begin{equation}	\label{mo.3'}
\varphi = B\psi
\iff
\Mat{\varphi} = \Mat{B}\Mat{\psi},
	\end{equation}
where $\Mat{\psi}\in M(n,1)$ is the matrix of the components of $\psi$
in the basis given. Comparing~\eref{mo.1} and~\eref{mo.3}, we get
	\begin{equation}	\label{mo.4}
B_{~\beta}^{\alpha}
= \sum_{\mu,\nu} \bigl( f^{-1}(x) \bigr)_{\mu}^{\alpha} \,
	b_{~\nu}^{\mu} \circ
	\bigl( f_{\beta}^{\nu}(x)\id_{\mathit{K}^1} \bigr) .
	\end{equation}

	For any matrix $C=[c_{~\beta}^{\alpha}]\in GL(n,\mathbb{C})$,
considered as a matrix operator (see above), we have
	\begin{equation}	\label{mo.5}
\Mat{C} = \bigl[C_{~\beta}^{\alpha}(x)\bigr],
\quad
C_{~\beta}^{\alpha}(x) =
\bigl( f^{-1}(x) \bigr)_{\mu}^{\alpha} \, c_{~\nu}^{\mu} f_{\beta}^{\nu},
\quad
C(f_\beta) = C_{~\beta}^{\alpha} f_\alpha .
	\end{equation}
Therefore, as one can expect, the matrix of the unit matrix
$\openone_n=\bigl[\delta_{\alpha}^{\beta}\bigr]_{\alpha,\beta=1}^{n}$
is exactly the unit matrix,
	\begin{equation}	\label{mo.6}
\pmb{\openone}_n = \openone_n .
	\end{equation}
If $\{f_\mu(x)\}$ does not depend on $x$, \ie if $f_{\alpha}^{\beta}$ are
complex constants, and $b_{~\nu}^{\mu}$ are linear, than~\eref{mo.4} implies
	\begin{equation}	\label{mo.7}
B_{~\beta}^{\alpha}
=
\bigl(f^{-1}(x)\bigr)_{\mu}^{\alpha} \, f_{\beta}^{\nu}(x) b_{~\nu}^{\mu}
\ \text{ for }\
\pd f_{\alpha}^{\beta}(x)/\pd x^\mu = 0 .
	\end{equation}
In particular, for a linear matrix operator $B$, we have
	\begin{equation}	\label{mo.8}
\Mat{B} = B \ \text{ for }\  f_{\beta}^{\alpha}
	= \delta_{\beta}^{\alpha}
	\ \text{ (\ie for $f_\alpha(x)=f_{\alpha}^{0}$)} .
	\end{equation}
Combining~\eref{mo.2} and~\eref{mo.4}, we deduce that the matrix of a product
of matrix operators is the product of the corresponding matrices:
	\begin{equation}	\label{mo.9}
C=A\odot B \quad\iff\quad
\Mat{C} = \Mat{A}\Mat{B} = \Mat{A}\odot\Mat{B}.
	\end{equation}

	After a simple calculation, we find the matrices of
$\openone_4\pd_\mu$ and $\slashed{\pd}=\gamma^\mu\pd_\mu$:%
\footnote{Here and below the Greek indices run from 0 to 3.}
	\begin{equation}	\label{mo.10}
\Mat{\pd}_\mu = \openone_4\pd_\mu + \ope{E}_\mu(x), \qquad
\Mat{\slashed{\pd}}
=
\ope{G}^\mu(x) \bigl( \openone_4\pd_\mu + \ope{E}_\mu(x) \bigr)
=
\ope{G}^\mu(x) \Mat{\pd}_\mu
	\end{equation}
where
\(
\ope{G}^\mu(x) = \bigl[ \ope{G}_{~\alpha}^{\lambda\mu}(x)
\bigr]_{\alpha,\lambda=0}^{3}
\)
and
\(
\ope{E}_\mu(x) = \bigl[ \ope{E}_{\mu\alpha}^{\lambda}(x)
\bigr]_{\alpha,\lambda=0}^{3}
\)
are defined via the expansions
	\begin{equation}	\label{mo.11}
\gamma^\mu f_\nu(x)	=: \ope{G}_{~\nu}^{\lambda\mu}(x) f_\lambda(x),
\quad
\pd_\mu f_\nu(x)	=: \ope{E}_{\mu\nu}^{\lambda}(x) f_\lambda(x) .
	\end{equation}
So, $\ope{G}^\mu$ is the matrix of $\gamma^\mu$ considered as a matrix
operator (see~\eref{mo.5}).%
\footnote{%
We denote the matrix of $\gamma^\mu$ by $\ope{G}^\mu$ instead of
$\boldsymbol{\gamma}^\mu(x)$ as usually~\cite{Bjorken&Drell-1,Itzykson&Zuber}
by $\boldsymbol{\gamma}$ is denoted the matrix 3\ndash vector
$(\gamma^1,\gamma^2,\gamma^3)$.%
}
Evidently
	\begin{equation}	\label{mo.12}
\Mat{\slashed{\pd}} = \slashed{\pd} \ \text{ for }\
	f_\alpha(x)=f_{\alpha}^{0} .
	\end{equation}

	Combining~\eref{mo.5},\eref{mo.6}, and~\eref{mo.10}, we get that the
matrix of the matrix operator $\ih\Slashed{D}-mc\openone_4$, entering in the
Dirac equation~\eref{7.1}, is
	\begin{equation}	\label{mo.13}
\ih\Mat{\Slashed{D}}-mc\pmb{\openone}_4 =
\ih\ope{G}^\mu(x)\bigl(\openone_4D_\mu + \ope{E}_\mu(x)\bigr) - mc\openone_4,
	\end{equation}
so that
	\begin{equation}	\label{mo.14}
\ih\Mat{\Slashed{D}}-mc\pmb{\openone}_4 =
\ih\Slashed{D}-mc\openone_4
\ \text{ for }\
f_\alpha(x) = f_{\alpha}^{0}.
	\end{equation}

\addcontentsline{toc}{section}{References}
\bibliography{bozhopub,bozhoref}
\bibliographystyle{unsrt}

\addcontentsline{toc}{subsubsection}{This article ends at page}

\addtocontents{toc}{\vspace{3ex}\noindent
Distribution of the material:\\[2ex]
I\ \ \ (1--3): Time-dependent approach,\ \ \
	arXiv E-print No.\ quant-ph/0105056 \\
II~(4\&5 \& A\&B): Covariant approach,\
	arXiv E-print No.\ quant-ph/0107002\\
}

\end{document}